\documentclass[sigconf, nonacm]{acmart}

\pdfoutput=1
\newcommand\vldbdoi{10.14778/3717755.3717774}
\newcommand\vldbpages{1169 - 1182}
\newcommand\vldbvolume{18}
\newcommand\vldbissue{4}
\newcommand\vldbyear{2024}
\newcommand\vldbauthors{\authors}
\newcommand\vldbtitle{\shorttitle} 
\newcommand\vldbavailabilityurl{URL_TO_YOUR_ARTIFACTS}
\newcommand\vldbpagestyle{empty} 

\usepackage{bbding}
\usepackage{subfigure} 
\usepackage[ruled,vlined]{algorithm2e}
\usepackage{graphicx} 
\usepackage{float}  
\usepackage{subfigure} 
\usepackage{multirow}
\usepackage[framemethod=tikz]{mdframed}

\usepackage{enumitem}

\begin{document}
\title{RCRank: Multimodal Ranking of Root Causes of Slow Queries\\ in Cloud Database Systems}

%%
%% The "author" command and its associated commands are used to define the authors and their affiliations.
\author{Biao Ouyang}
\affiliation{%
  \institution{East China Normal University, China}
  \country{}
  }
\email{bouyang@stu.ecnu.edu.cn}

\author{Yingying Zhang}
\affiliation{%
   \institution{Alibaba Cloud Computing, China}
  \country{}
  }
\email{congrong.zyy@alibaba-inc.com}
\author{Hanyin Cheng}
\affiliation{%
   \institution{East China Normal University, China}
  \country{}
  }
\email{hanyinch@gmail.com}
\author{Yang Shu}
\authornote{Corresponding author}
\affiliation{%
   \institution{East China Normal University, China}
  \country{}
  }
\email{yshu@dase.ecnu.edu.cn}
\author{Chenjuan Guo}
\affiliation{%
  \institution{East China Normal University, China}
  \country{}
  }
\email{cjguo@dase.ecnu.edu.cn}
\author{Bin Yang}
\affiliation{%
  \institution{East China Normal University, China}
  \country{} }
\email{byang@dase.ecnu.edu.cn}
\author{Qingsong Wen}
\affiliation{%
  \institution{Squirrel Ai Learning, USA}
  \country{}}
\email{qingsongedu@gmail.com}
\author{Lunting Fan}
\affiliation{%
  \institution{Alibaba Cloud Computing, China}
  \country{}}
\email{lunting.fan@taobao.com}

\author{Christian S. Jensen}
\affiliation{%
  \institution{Aalborg University, Denmark}
  \country{}}
\email{csj@cs.aau.dk}

\begin{abstract}
With the continued migration of storage to cloud database systems, the impact of slow queries in such systems on services and user experience is increasing. Root-cause diagnosis plays an indispensable role in facilitating slow-query detection and revision. 
This paper proposes a method capable of both identifying possible root cause types for slow queries and ranking these according to their potential for accelerating slow queries. 
This enables prioritizing root causes with the highest impact, in turn improving slow-query revision effectiveness.
To enable more accurate and detailed diagnoses, we propose the multimodal Ranking for the Root Causes of slow queries (\texttt{RCRank}) framework, which formulates root cause analysis as a multimodal machine learning problem and 
leverages multimodal information from query statements, execution plans, execution logs, and key performance indicators. 
To obtain expressive embeddings from its heterogeneous multimodal input, \texttt{RCRank} integrates self-supervised pre-training that enhances cross-modal alignment and task relevance. Next, the framework integrates root-cause-adaptive cross Transformers that enable adaptive fusion of multimodal features 
with varying characteristics. Finally, the framework offers a unified model that features an impact-aware training objective for identifying and ranking root causes. We report on experiments on real and synthetic datasets, finding that \texttt{RCRank} is capable of consistently outperforming the state-of-the-art methods at root cause identification and ranking according to a range of metrics.
  
\end{abstract}

\maketitle

\pagestyle{\vldbpagestyle}
\begingroup\small\noindent\raggedright\textbf{PVLDB Reference Format:}\\
\vldbauthors. \vldbtitle. PVLDB, \vldbvolume(\vldbissue): \vldbpages, \vldbyear.\\
\href{https://doi.org/\vldbdoi}{doi:\vldbdoi}
\endgroup
\begingroup
\renewcommand\thefootnote{}\footnote{\noindent
This work is licensed under the Creative Commons BY-NC-ND 4.0 International License. Visit \url{https://creativecommons.org/licenses/by-nc-nd/4.0/} to view a copy of this license. For any use beyond those covered by this license, obtain permission by emailing \href{mailto:info@vldb.org}{info@vldb.org}. Copyright is held by the owner/author(s). Publication rights licensed to the VLDB Endowment. \\
\raggedright Proceedings of the VLDB Endowment, Vol. \vldbvolume, No. \vldbissue\ %
ISSN 2150-8097. \\
\href{https://doi.org/\vldbdoi}{doi:\vldbdoi} \\
}\addtocounter{footnote}{-1}\endgroup

\ifdefempty{\vldbavailabilityurl}{}{
\vspace{.3cm}
\begingroup\small\noindent\raggedright\textbf{PVLDB Artifact Availability:}\\
The source code, data, and/or other artifacts have been made available at \url{https://github.com/decisionintelligence/RCRank}.
\endgroup
}

\section{Introduction}
\label{sec:intro}
Companies and individuals are increasingly migrating their database services to the cloud. At the same time, poor cloud database system performance in the form of slow queries causes economic losses to users and decreases user trust in cloud-based data management. For example, every 0.1-second increase in Amazon's page load time causes a 1\% drop in sales, and a 0.5-second increase in Google's search latency causes a 20\% drop in traffic~\cite{RN16}.
Thus, accelerating slow queries is essential for ensuring high-performance cloud database systems.

Slow queries may be attributed to database system internal factors, such as the absence of relevant indexes or poorly written SQL statements, and external factors, such as I/O bottlenecks and network issues. In this paper, we aim at providing a framework to enable users to address slow queries while focusing on root causes deriving from internal factors. This is because external factors are often beyond the control of users and require the intervention of cloud database service providers. In contrast, internal factors can be controlled by cloud database users, such as by modifying indexes or rewriting queries.

\begin{figure}[t] 
	\centering 
	\subfigure[Root Cause Identification (RCI)]{
		\label{fig:identify}
		\includegraphics[width=0.9\linewidth]{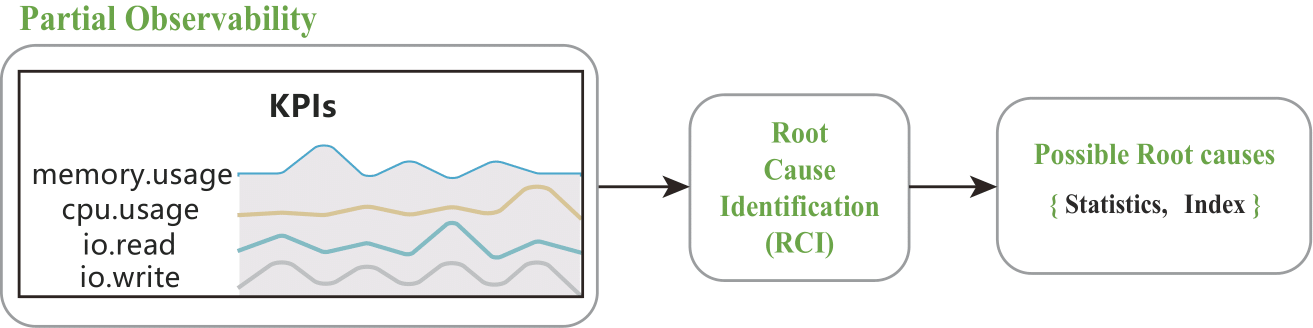}}

	\subfigure[Root Cause Ranking (RCR)]{
		\label{fig:rank}
		\includegraphics[width=0.9\linewidth]{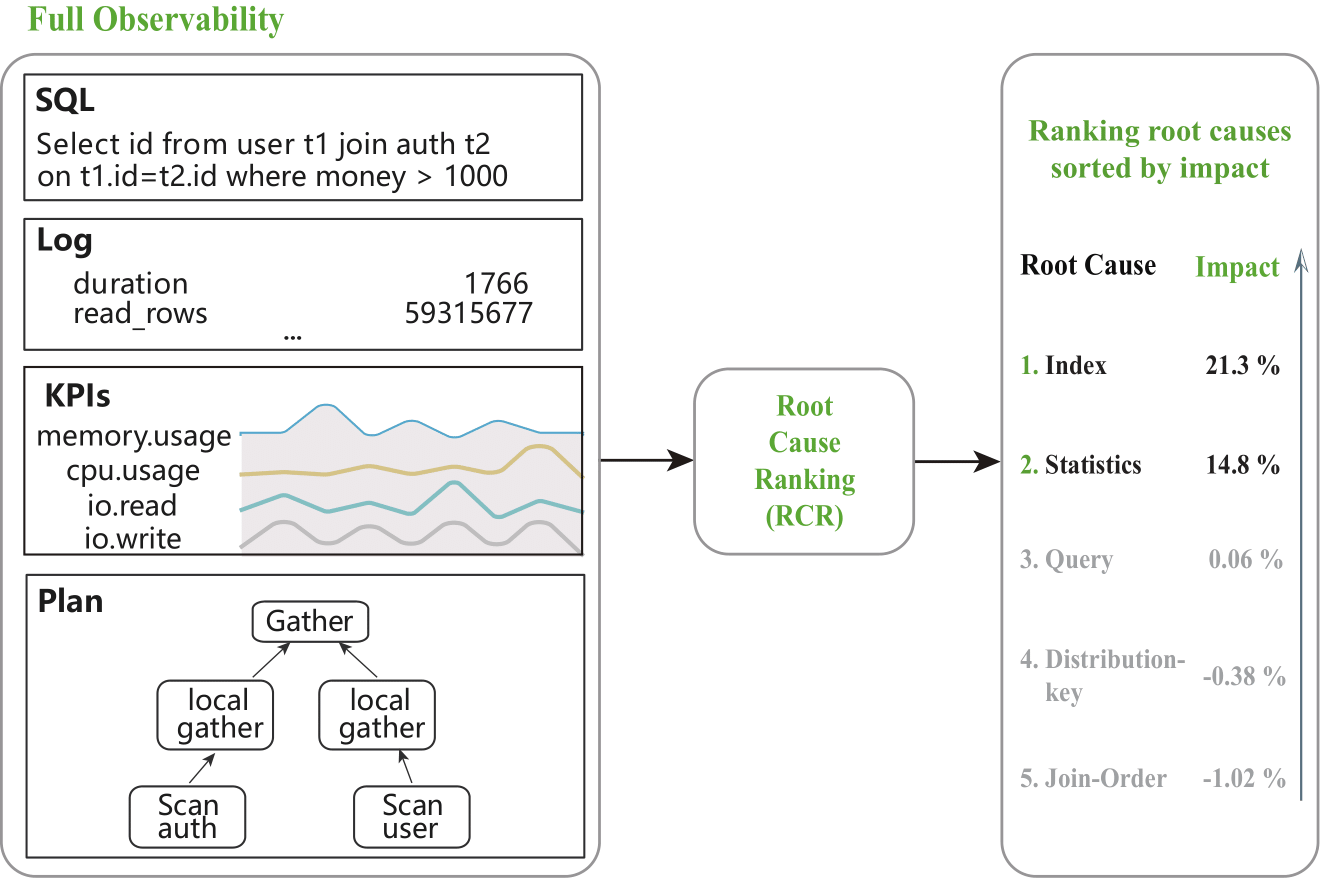}}  
	\caption{Root cause identification (RCI) vs. root cause ranking (RCR). (i) RCI often utilizes partial observability, whereas RCR utilizes multimodal, full observability. (ii) RCI only identifies possible root causes, whereas RCR ranks root causes according to their potential impact, enabling users to identify the most significant root causes. }
	\label{fig:rank_identify}
\end{figure}

Identifying root causes, i.e., factors that cause slow queries, allows for performance improvement through corresponding revision methods. Thus, root cause analysis identifies the causes of slow-query types through root cause classification~\cite{DBLP:journals/pvldb/iSQUAD, DBLP:journals/corr/DBot, DBLP:journals/pvldb/DBmind}.  
While methods exist that target slow-query identification, two limitations remain:\\
\textbf{Identification of Root Causes Mainly.} 
Existing methods~\cite{DBLP:journals/corr/DBot,DBLP:journals/pvldb/iSQUAD,DBLP:conf/sigmod/DBSherlock} focus on identifying root causes, as shown in Figure~\ref{fig:rank_identify}. This does not fully support the need to focus on addressing the ``most significant'' root causes.  
Revising slow queries based on root causes is costly, and making revisions according to every root cause can result in 
substantial expenses. Therefore, it is attractive to take into account the impact of root causes (i.e., how much runtime can be 
saved if a root cause is addressed) when choosing which one to address. 
Root Cause Identification (RCI) fails to quantify the potential impact of addressing identified root causes, thus limiting the practical utility. 

\noindent
\textbf{Incomplete Observability of Cloud Database Systems. } 
As shown in Figure~\ref{fig:rank_identify}, most methods~\cite{DBLP:journals/pvldb/iSQUAD,DBLP:conf/sigmod/DBSherlock} rely on single-modal information, e.g., key performance indicators, such as CPU or memory use time series, to identify root causes, disregarding other data sources that can offer insight into internal factors related to slow query processing. For example, query statements and execution plans contain information about query objectives and estimated execution processes; and execution logs contain information about resources consumed during query execution and execution status. 
Aiming for full observability by taking all these data sources into account offers a better foundation for understanding slow queries and their causes, thereby enabling more accurate root cause identification.  

To eliminate the two limitations, we propose ranking the impact of the root causes using multimodal observability data.

\textbf{Ranking the impact of root causes. } We propose a model to estimate the impact of root causes, thus providing a ranked list of identified root causes. To unify identification and ranking, we propose a training objective for ranking root causes by estimating their impact. Two impact-aware regularizations are utilized in the training objective to enhance the identification of valid root causes and preserve the order of their impacts. Specifically, we collect slow queries from the cloud database system that encompass query statements, execution plans, execution logs, and key performance indicators. We then revise the queries according to potential root causes and record the relative improvement of the revised queries as the impact of each root cause. For instance, the runtime of a given slow query may drop by 14.8\%, after updating Outdated Statistics. Then we associate 14.8\% with the ``Outdated Statistics'' root cause for the query. The multimodal information about queries, along with the impact of root causes are used for training a model. During inference, the model (i) diagnoses slow queries to identify root causes that lead to the slow queries and (ii) ranks the root causes according to their potential impact.

\textbf{Encoding multimodal representations. } We encode the available multimodal information into a unified embedding to support root cause ranking. The information encompasses query statements in text form, execution plans represented as graphs, logs in key-value format, and key performance indicators (KPIs) as time series. To integrate this information, we first use different encoders to encode each modality into embeddings. Then we employ a fusion embedding module to encode the heterogeneous embeddings of the different modalities into a shared embedding space for further feature extraction. 
We propose self-supervised pre-training to enhance cross-modal alignment and task relevance to obtain more expressive embeddings. To enable effective fusion of multimodal features, we propose a multimodal fusion model that uses root-cause-adaptive cross Transformers, capable of adaptively fusing multimodal features for the diagnosis of 
root causes with varying characteristics. 

To the best of our knowledge, this is the first multimodal method for ranking the root causes of slow queries for root cause diagnosis. 
The contributions are summarized as follows.
\begin{itemize}[leftmargin=*]
    \item We propose means to automatically collect root cause impact labeled data, which provides a data foundation for multimodal root cause diagnosis.
    \item We propose a multimodal representation method that is capable of harnessing full observability by implementing expressions for embedding heterogeneous modalities, modality alignment, and task relevance, while effectively and adaptively fusing multimodal features.
    \item We propose a root cause impact ranking method that enables unified identification and ranking of root causes. 
    \item We consider datasets from the Alibaba Hologres system as well as synthetic datasets. Experimental results show that our method is capable of outperforming existing approaches in terms of root cause identification and ranking according to a range of metrics.
\end{itemize}

\section{RELATED WORK}
\label{sec:releted}
\subsection{Slow Query Analyses}
\label{sec:slowquery_releted}
We classify slow-query analysis methods in Table~\ref{table:RelatedWork} from two perspectives: root cause identification vs. ranking for slow queries and single-modality vs. multi-modality modeling. 

Existing Root Cause Identification (\textbf{RCI}) methods~\cite{DBLP:journals/pvldb/iSQUAD, DBLP:conf/sigmod/ExplainIt!, DBLP:conf/sigmod/DBSherlock, DBLP:journals/pvldb/Opengauss, DBLP:journals/corr/DBot} use clustering, attribution, or prediction methods to identify root causes. However, these methods cannot estimate or rank the impact of root causes (\textbf{RCR}). For example, when diagnosing root causes of slow queries, multiple root causes can contribute to an issue, so that addressing each can lead to different levels of performance improvement. Users are interested primarily in root causes that can bring the most significant performance gains.

Previous studies primarily focus on slow-query analysis through \textbf{single-modal input}, e.g., key performance indicators. 
DBSherlock~\cite{DBLP:conf/sigmod/DBSherlock} focuses on the precise values of KPIs. A specific KPI value can only suggest whether or not an indicator is anomalous. iSQUAD~\cite{DBLP:journals/pvldb/iSQUAD} focuses on the root causes of intermittently slow queries and can effectively inform DBAs of the specific category of an anomaly. However, when iSQUAD encounters root causes outside its clustering, manual analysis of the anomaly types of KPIs is required. ExplainIt!~\cite{DBLP:conf/sigmod/ExplainIt!} employs a probabilistic graphical model facilitating causal reasoning to analyze root causes. 
Interpretable ML methods~\cite{DBLP:conf/sigmod/DBSherlock, DBLP:journals/pvldb/iSQUAD, DBLP:journals/tcc/Dundjerski} focus on root cause analysis through KPIs and offer reasonable interpretations. 

Two existing methods fuse \textbf{multimodal input} encompassing system metrics and query information to identify root causes. OpenGauss~\cite{DBLP:journals/pvldb/Opengauss} discovers time-consuming operators in SQL by fusing execution plans and system KPI metrics. D-Bot~\cite{DBLP:journals/corr/DBot} leverages large language models~\cite{DBLP:conf/nips/instructions, DBLP:conf/nips/CoT, DBLP:conf/nips/ToT,zhu2024chat2query,zhu2023autotqa} alongside maintenance knowledge to analyze monitoring and query information, facilitating the identification of root causes of slow queries. However, in our scenario, additional observable information sources are available. 
Since users prioritize addressing root causes of slow queries with high impact, we not only identify the root causes but also rank these according to their impact to facilitate effective slow-query resolution. 
Our study considers query statements, execution plans, execution logs, and time series to rank root causes of slow queries according to their impact.

\begin{table}[t]
\centering
\caption{Categorization of slow query diagnosis.}
\resizebox{0.4\textwidth}{!}{
\begin{tabular}{c|c|c} 
\toprule
    & 
   \begin{tabular}[c]{@{}c@{}}
   \textbf{Root Cause} \\ 
   \textbf{Identification}
   \textbf{(RCI)}
   \end{tabular}
   & 
   \begin{tabular}[c]{@{}c@{}}
   \textbf{Root Cause} \\ 
   \textbf{Ranking} 
   \textbf{(RCR)}
   \end{tabular}
   \\ 
\hline
\begin{tabular}[c]{@{}l@{}}
\textbf{Single-Modality} \\ \textbf{Observability}
\end{tabular} 
&     \begin{tabular}[c]{@{}l@{}}
iSQUAD~\cite{DBLP:journals/pvldb/iSQUAD}   \\ 
ExplainIt!~\cite{DBLP:conf/sigmod/ExplainIt!}   \\
DBSherlock~\cite{DBLP:conf/sigmod/DBSherlock}   
\end{tabular}          &         -       \\ 
\hline
\begin{tabular}[c]{@{}l@{}}
\textbf{Multi-Modality} \\ \textbf{Observability}
\end{tabular} &   \begin{tabular}[c]{@{}l@{}}OpenGauss~\cite{DBLP:journals/pvldb/Opengauss} \\ D-BOT~\cite{DBLP:journals/corr/DBot} \end{tabular}     &     \texttt{RCRank} (ours)          \\
\bottomrule
\end{tabular}

\label{table:RelatedWork}
}
\end{table}

\subsection{Multimodal Learning}
Previous multimodal learning methods~\cite{DBLP:conf/iccv/VideoBERT,DBLP:conf/icml/FusedAcoustic} focus on the fusion of text, images, and audio. The fusion typically involves feature fusion and alignment through concatenation and cross-attention. Tokens from multiple modalities are concatenated directly, adding positional information to distinguish the modality of each input. The aim is to understand the context of other modalities. However, this approach comes with high computational complexity. Instead, we use the cross-attention mechanism to reduce computational complexity. Cross-Attention~\cite{DBLP:conf/nips/ViLBERT, DBLP:conf/icml/BLIP, DBLP:conf/nips/ALBEF,DBLP:conf/eccv/LS, DBLP:conf/kdd/Grained} exchanges information between modalities. However, the large number of modalities may still result in an increase in computational complexity when pairwise calculations are involved. We thus use the primary modality to avoid pairwise interactions between all modalities.

When fusing text, images, and audio, existing methods extract commonalities and specificities across the modalities~\cite{DBLP:conf/aaai/TAILOR, DBLP:conf/mm/MISA, DBLP:conf/aaai/sc1}. The methods achieve this by using shared modules for extracting common features across text, images, and audio, while using individual modules to capture modality-specific characteristics. 
AnoFusion~\cite{DBLP:conf/kdd/RobustMicroservice} models the failure instance detection problem by capturing the relationship between time series, graphs, and text. Our method focuses on the complex correlations among text, graphs, and time series~\cite{DBLP:journals/pacmmod/0002Z0KGJ23,DBLP:journals/pacmmod/Wu0ZG0J23,DBLP:journals/pvldb/QiuHZWDZGZJSY24,DBLP:journals/vldb/WuWYZGQHSJ24,DBLP:conf/iclr/ChenZ0SWW0G24,DBLP:journals/pvldb/ZhaoGCHZY23,DBLP:journals/pvldb/ChengCGZWYJ23,tian2024air} to rank the root causes of slow queries.

Additionally, some methods focus on improving query performance, with deep learning representation methods pre-training on SQL and query plans~\cite{DBLP:conf/sigmod/PreQR, DBLP:journals/pvldb/Queryformer}, and encoding plan methods~\cite{DBLP:journals/pvldb/Queryformer,DBLP:conf/sigmod/Plan1, DBLP:conf/icde/Plan2,DBLP:journals/pvldb/plan3}. Some methods encode SQL for cost estimation~\cite{DBLP:conf/sigmod/PreQR, DBLP:conf/cidr/SQL1, DBLP:journals/pvldb/SQL2} and query optimization~\cite{DBLP:conf/cidr/opti25, DBLP:journals/pvldb/opti01, DBLP:conf/sigmod/opti2, DBLP:conf/sigmod/tune42, DBLP:conf/sigmod/dbopt1}, and some method~\cite{DBLP:journals/pvldb/PanWZY0CGWTDZYZ23} encodes metrics for autoscaling. Some methods focus on enhancing query performance through the automatic adjustment of indexes~\cite{DBLP:conf/sigmod/indexC13, DBLP:journals/pvldb/index028, DBLP:journals/pvldb/indexP40, DBLP:conf/sigmod/indexN19, DBLP:conf/icde/Extend}. In contrast, we utilize more extensive information for identifying and ranking root causes of slow queries.

\section{Problem Setting and Formulation}

\subsection{Query Formalization}

A query $X_i=(S, P, L, I)$ is a four-tuple composed of a query statement $S$, an execution plan $P$, an execution log $L$, and KPIs $I$. 
We define this as follows.

\begin{itemize}[leftmargin=*]
\item {\verb|Query Statement (SQL)|}: 
A query statement is the SQL statement of the query, which is a sequence $S= \langle s_1, s_2, ...,s_n \rangle $, where $s_i$ is an SQL text token (e.g., ``Select'').
This captures the intent of the query and relevant database information, e.g., 
table names.
\item {\verb|Execution Plan (Plan)|}:
An execution plan is a Directed Acyclic Graph (DAG) $P=(V, E)$, where $V$ is a set of operators and a directed edge $e_k$ from $v_i$ to $v_j$ exists if operator $v_j$ is the next operation of $v_i$.
\item {\verb|Execution Log (Log)|}: Log files record information related to queries, e.g., the query duration and number of rows read by the execution of the query. In our setting, logs are represented as vectors $L=  ( l_1, l_2, ..., l_m )$, where $l_i$ is a record related to the query.
\item {\verb|Key Performance Indicators (KPIs)|}: The execution 
times of query statements are influenced by the current performance state of the database. KPIs $I =  ( I_1, I_2, ..., I_q ) $ are composed of $q$ multi-dimensional time series, each of length t, that are employed for evaluating database performance. Here $q$ is the number of evaluated metrics. These include memory usage percentages, CPU utilization percentages, and I/O counts.
\end{itemize}

\subsection{Root Causes for Slow Queries}
\label{sec:rc_sq}

Many factors can lead to slow queries, including external factors such as CPU and memory depletion, resource contention leading to deadlocks, and instance node crashes. Internal factors such as sub-optimal database structures and poorly-written query statements also contribute to slow queries. While extensive research exists on diagnosing the root causes of slow queries related to external factors, database users are often interested primarily in the revisions they can make in terms of database structures and query statements to speed up query execution.

We define a \textbf{root cause} as an internal factor within the database, e.g., a slow query caused by sub-optimal database structures or poorly-written query statements. For example, the root cause of \textit{missing indexes} means that, in scenarios with a large volume of data, queries without indexes result in full table scans, consuming substantial computational resources and time. Adding appropriate indexes can enhance query efficiency. By employing suitable solutions to address root cause issues, we can improve the execution time of slow queries. 
After revising slow queries using revision methods corresponding to different root causes, the query performance is improved. Users are particularly concerned with root causes whose revision methods can contribute the most to improving query performance, e.g., the root cause that, when addressed, improve query execution time the most.

\textbf{The impact of a root cause} captures how much a slow query is influenced by the root cause. More precisely, consider a slow query $X_i$ and potential root causes $RC=\{RC_1, RC_2, ..., RC_r\}$, where $r$ is the total number of root cause types. We define an impact value $y_{ij}$ that captures how much the performance of the query can be improved when revised according to root cause $RC_j$:
\begin{equation}
\begin{aligned}
    y_{ij} = \frac{\mathit{runtime}(X_i)-\mathit{runtime}(\mathit{revise}(X_i, RC_j))}{\mathit{runtime}(X_i)}
\label{eq:impact}
\end{aligned}
\end{equation}
Here, $\mathit{revise}(X_i, RC_j)$ denotes the original slow query $X_i$ revised according to root cause $RC_j$. 
As an example, the original runtime of $X_i$ is 1.5 seconds, and $RC_j$ indicates missing indexes. Then $\mathit{revise}(X_i, RC_j)$ involves creating relevant indices for query $X_i$, and the revised runtime $runtime(revise(X_i,RC_j)$ becomes 0.67 seconds. Therefore, the impact $y_{ij}$ of $RC_j$ on $X_i$, is $(1.5-0.67)/1.5=55.3\%$, meaning that after revision w.r.t. root cause $RC_j$, the runtime of the slow query $X_i$ is improved by 55.3\%.

\textbf{Valid root causes} A root cause $RC_j$ is valid for slow query $X_i$ if $y_{ij}$ exceeds a threshold $\epsilon$. 

\subsection{Problem Formulation}
Given a root cause impact dataset $D = \{(y_{ij}, X_i, RC_j)\}$ consisting of impacts of root causes for different slow queries, we aim at learning a model $f$ that can estimate the impact $\hat{y}_{kj}$ for an unseen query $X_k$ for each root cause $RC_j$. This can be formulated as follows. 
\begin{equation}
\begin{aligned}
    \hat{y}_{kj} = f_{\theta}(X_k, RC_{j}),
\end{aligned}
\end{equation}
where $\theta$ represents the learnable parameters for the estimation model $f$ and $1 \leq j \leq r$.

\section{Multimodal Diagnosis Framework}
\begin{figure}[t]
\begin{center}
  \includegraphics[width=0.9\linewidth]{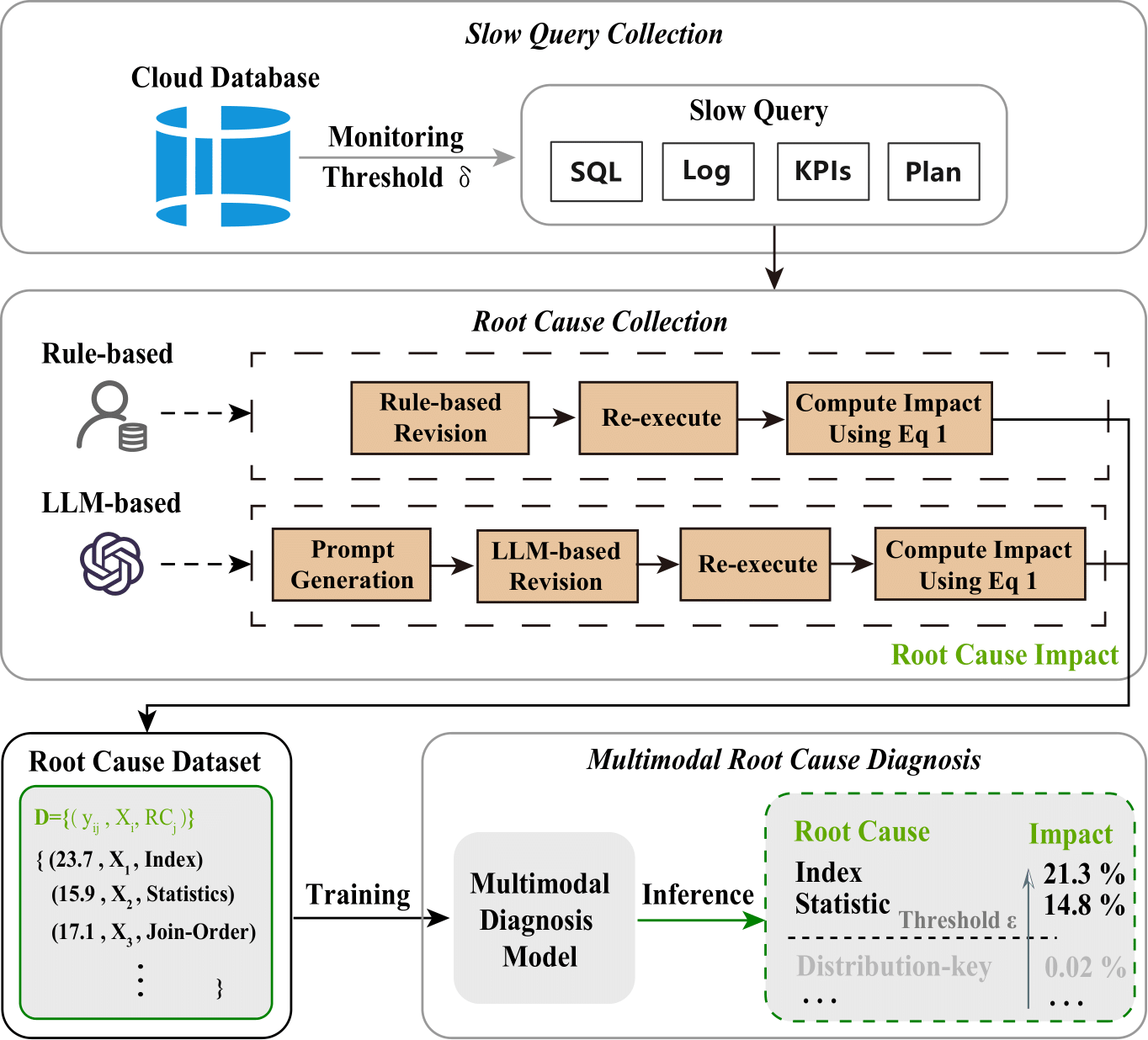} 
\end{center}
  \caption{Overview of the multimodal diagnosis framework for root causes of slow queries.}
  \label{fig:dataflow}
\end{figure}

\subsection{Overview}
We propose a multimodal diagnosis framework for identifying root causes of slow queries and ranking them by their impact. As shown in Figure~\ref{fig:dataflow}, the framework consists of two main modules: a slow query and root cause collection module, and a multimodal root cause diagnosis module.

The \textbf{slow query and root cause collection module} includes cloud database system monitoring, slow query collection, and root cause collection. The module collects slow queries and the impacts of root causes on the slow queries, which serves as a data foundation for the second module. The \textbf{multimodal root cause diagnosis module} learns a mapping from unseen slow queries $X_k$ and root causes $RC_j$ to the estimated impact $\hat{y}_{kj}$ of $RC_j$ on $X_k$, thus facilitating the construction of lists of root causes according to their estimated impact.

\subsection{Slow Query and Root Cause Collection}\label{sec:collection}

We collect slow queries and the impact of their corresponding root causes as $\{(y_{ij}, X_i, RC_j)\}$. 
As shown in Figure~\ref{fig:dataflow}, slow query collection involves continuous monitoring of queries and database instances through a cloud database monitoring system, collecting queries that exceed the slow query threshold $\delta$. Root cause collection involves obtaining slow query root causes through analysis by a rule-based method and a large language model (LLM) based method.

\textbf{Slow query collection} continuously monitors the execution times of queries through the cloud database monitoring system, collecting the query statements (\verb|SQL|), execution plans (\verb|Plan|), and execution logs (\verb|Log|) of the queries. The performance monitoring system continually gathers instance metric information (\verb|KPIs|), such as the memory usage percentage and CPU utilization percentage, thereby obtaining the status. We collect all monitored queries $\{Q_i\}$ for pre-training, with queries exceeding the slow query threshold $\delta$ considered as slow queries.

\textbf{Root cause collection} involves analyzing slow queries and revising slow queries according to different root causes. Therefore, we design two types of automated methods: \textit{rule-based} and \textit{LLM-based methods}. For root causes that can be addressed easily by employing predefined SQL templates, we choose rule-based methods. For the remaining cases, we use LLM-based methods.

\textit{Rule-based methods.} As shown in Figure~\ref{fig:dataflow}, the rule-based method includes three steps. First, according to a specific root cause $RC_j$, we employ corresponding rules to revise the slow query using predefined SQL templates. Second, we re-execute the slow query $X_i$ to obtain the execution time after revision. Third, we calculate the impact $y_{ij}$ according to Equation~\ref{eq:impact}. We detail the rule-based methods for two specific root causes for illustration.

$\star$ {Statistics}. Unsynchronized statistical information can result in the generation of sub-optimal execution plans, thus causing slow queries. For a slow query, we extract all tables involved in the query and then update the statistics by executing the SQL ``ANALYZE tablename'' for these tables. After updating the statistics, we re-execute the query and use Equation~\ref{eq:impact} to compute the impact of the root cause ``Statistics.''

$\star$ Join-Order. When SQL join relationships are complex or involve many tables, the optimizer spends considerable time selecting the optimal join order. To collect slow queries that require adjusted join strategies, we re-execute the slow queries by modifying the execution strategy. For example, we can utilize ``set optimizer\_join\_order = greedy;'' in SQL to adjust the execution strategy. Alternatively, we can use the ``exhaustive'' or ``query'' strategies. We record the runtime of the best join order, i.e., the join order with the shortest runtime. Based on this, we then compute the impact of the root cause ``Join-Order.'' 

\begin{figure}[t]
\begin{center}
  \includegraphics[width=0.9\linewidth]{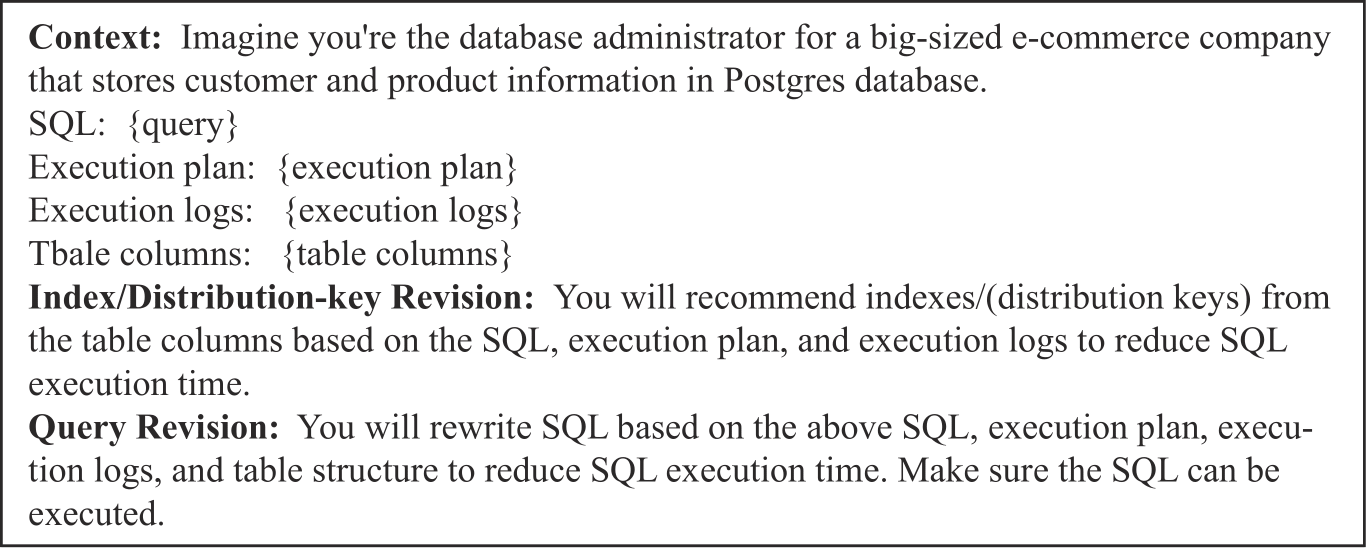}
\end{center}
  \caption{ 
  Prompt template with context and revision sections for LLM-based slow-query revision.}
\label{fig:revise}
\end{figure}

\textit{LLM-based methods.} As shown in Figure~\ref{fig:dataflow}, the LLM-based method encompasses four steps. First, we input query information, database information, and root cause types, and then we generate a prompt~\cite{DBLP:journals/csur/PromptServey} based on the prompt template shown in Figure~\ref{fig:revise}. Second, based on the generated prompt, the LLM recommends slow-query revisions. Third, we re-execute the revised query to obtain the execution time after revising the query according to root cause $RC_j$. Fourth, we calculate the impact $y_{ij}$ according to Equation~\ref{eq:impact}. The three root causes addressed using LLM-based methods are handled as follows.

$\star$ Index. With no indexes available, the database stem executes queries by scanning entire tables row by row to find matching rows. On the other hand, too many indexes can result in index inefficiency. We employ the LLM approach to recommend indexes that yield the shortest execution time. First, we provide the original query statement, execution plan, execution logs, and table structure information as a prompt; see Figure~\ref{fig:revise}. Then, the LLM recommends indexes based on the prompt. We compare the execution times when using the indexes recommended by the LLM and those when using the original indexes to determine the impact of the root cause ``Index.''

$\star$ Distribution-key. The distribution key facilitates the partitioning of data into shards, and a uniform distribution helps prevent data skew.  We provide the original query statement, execution plan, execution logs, and table structure information as a prompt; again, see  Figure~\ref{fig:revise}. We compare the execution time of the distribution keys recommended by the LLM with those of the original distribution keys to determine the impact of the root cause ``Distribution-key.''  

$\star$ Query. Queries written by users may contain redundant operators, nested subqueries, etc., so that reduced execution times can result from rewriting the queries into equivalent queries. We provide the original query statement, execution plan, execution logs, and table structure information as a prompt; again, see Figure~\ref{fig:revise}. The LLM recommends rewritten queries based on the prompts. Considering that the LLM may output incorrect queries, we inform the LLM about erroneous queries along with the corresponding error messages to get better recommendations. We compare the execution times of rewritten queries with those of the original query to determine the impact of the root cause ``Query.''

By collecting slow queries and revising them according to methods associated with corresponding root causes, we construct a {dataset} $D = \{(y_{ij}; X_i, RC_j)\}$ for identifying and ranking root causes of slow queries.

\subsection{Multimodal Root Cause Diagnosis Model}
\begin{figure*}[t]

\begin{center}
  \includegraphics[width=0.9\textwidth]{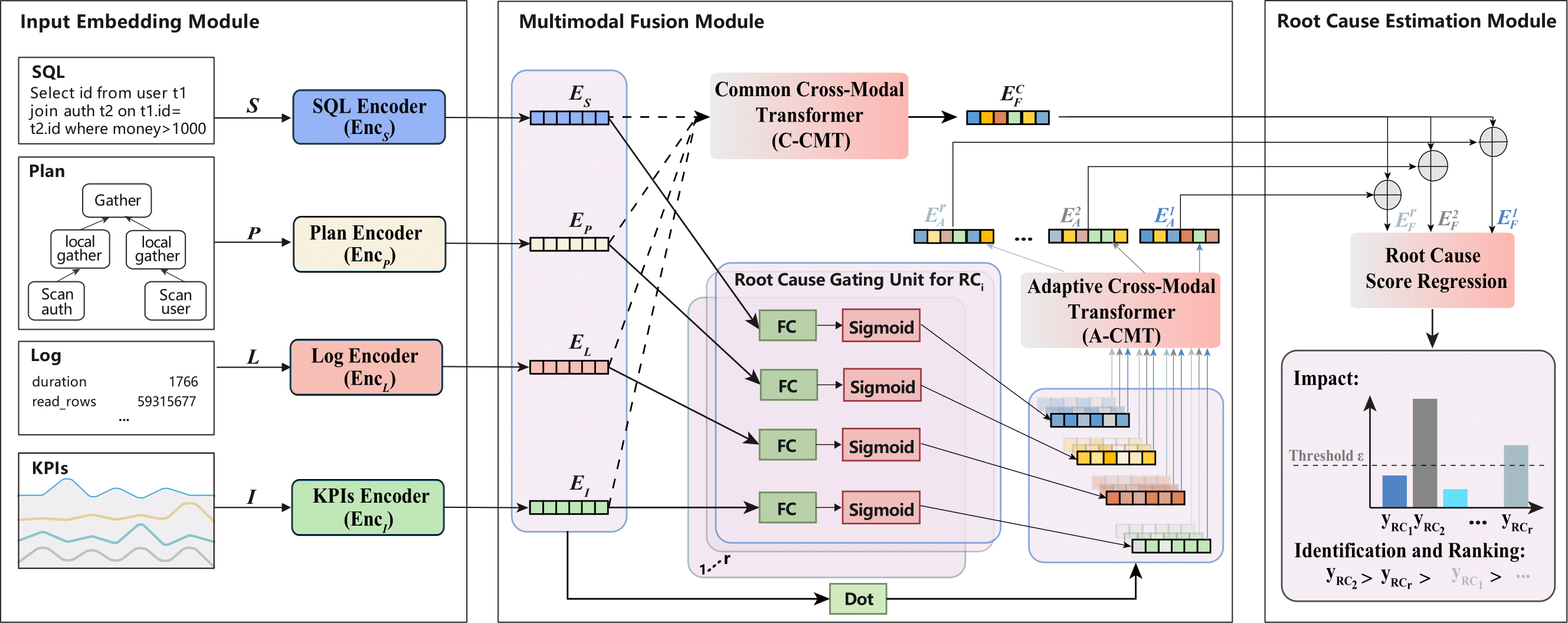}
\end{center}
  \caption{Overview of the multimodal learning model for root cause diagnosis, which is composed of three main modules: (1) input embedding module, (2) multimodal fusion module, and (3) root cause estimation module.}
  \label{fig:Overview}
\end{figure*}

We propose a multimodal diagnosis model for identifying the root causes of slow queries and ranking the impact of different root causes based on dataset $D = \{(y_{ij}, X_i, RC_j)\}$. A overview of the framework and its three main modules is shown in Figure~\ref{fig:Overview}.

The \textbf{input embedding} module incorporates an encoder for each input modality. Given the multimodal input data $X_i$ consisting of a query statement $S$, an execution plan $P$, an execution log $L$, and key performance indicators $I$, the encoders output the embeddings of each modality $\mathbb{E}(X_i)=(E_S, E_P, E_L, E_I)$. The module determines how we encode the heterogeneous inputs of different modalities into a shared embedding space for further feature extraction. Note that the proposed pre-training processes are self-supervised~\cite{DBLP:conf/ijcai/YangGHT021}, which reduces the amount of root-cause-annotated data needed for supervised training and enables the utilization of unlabeled query data for obtaining more expressive embeddings. In Section~\ref{sec:input embedding}, we describe the design of the input embedding module. To improve the quality of embeddings for better cross-modal alignment and task relevance and to reduce the amount of root-cause-annotated data needed for training, we employ self-supervised multimodal pre-training for the input embeddings.

The \textbf{multimodal fusion} module takes the encoded embeddings of each modality as input. The module learns to fuse information from these embeddings to extract the features $E_F=f_\text{fusion}(E_S, E_P, E_L, E_I)$ for root cause diagnosis. To facilitate feature extraction in the module, we propose a cross-modal Transformer architecture in Section~\ref{sec:multimodal fusion}. We design feature decomposition and adaptive gating mechanisms to enable adaptive multimodal fusion for different root causes.

The \textbf{root cause estimation} module leverages the extracted multimodal fusion features to obtain the final estimations $Y_{RC}=f_\text{pred}(E_F)$ regarding the root causes. This module's training and inference processes are designed to identify and rank the impacts of root causes. It includes a learning objective for estimating the impacts of different root causes to identify and rank root causes in a unified model. The estimation accuracy is enhanced by impact-aware regularization for valid and highly impactful root causes. This is covered in Section~\ref{sec:root cause prediction}.

\subsubsection{Self-supervised Pre-trained Embeddings}\label{sec:input embedding} 
\ 
\newline \textbf{Input Embeddings.} The input embedding module encodes inputs with different modalities. Given the multimodal input data $X_i$, we employ different encoders tailored for the different modalities in the input (denoted as $\text{Enc}$) to obtain an embedding for each single modality in a common embedding space:
\begin{equation}
\begin{aligned}
    (E_S, E_P, E_L, E_I) &= \mathbb{E}_{\phi}(X_i) \\
    &= \left(\text{Enc}_S(S), \text{Enc}_P(P), \text{Enc}_L(L), \text{Enc}_{I}(I)\right).
\end{aligned}
\end{equation}
More specifically, we employ BERT~\cite{DBLP:conf/naacl/Bert} as the encoder of query statements $S$, QueryFormer~\cite{DBLP:journals/pvldb/Queryformer} as the encoder of execution plans $P$, a multi-layer perceptron as the encoder of execution logs $L$, and a 2DCNN model as the encoder of KPIs $I$.

\textbf{Self-supervised Multimodal Pre-training.} 
In the multimodal data inputs, partial correspondence exists among the different modalities. For example, field names, table names, and filter conditions in query statements also occur in nodes in the execution plan tree. Likewise, some data in execution logs has corresponding data in the root nodes of plans. Inspired by such correspondences in the multimodal query data, we propose self-supervised pre-training to align the different modalities and enhance the expressiveness of the data embeddings.

\begin{figure}[t]
\begin{center}

  \includegraphics[width=0.9\linewidth]{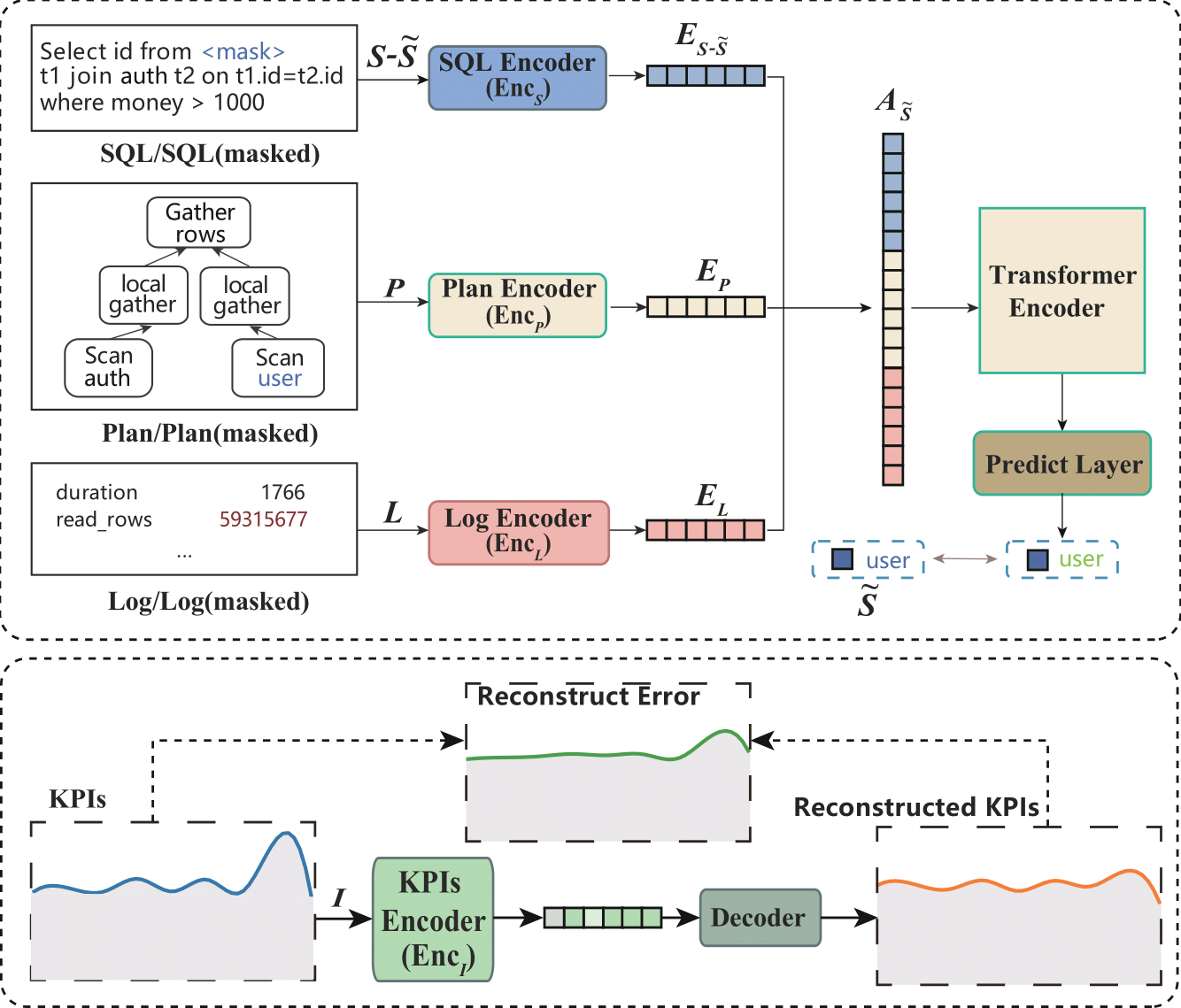}
\end{center}
  \caption{Illustration of self-supervised pre-training for multimodal encoders $Enc_S$, $Enc_Q$, $Enc_L$, and $Enc_I$.}
  \label{fig:pretrain}
\end{figure}

As shown in Figure~\ref{fig:pretrain}, to align query statements, execution plans, and execution logs, we propose to mask critical information within each modality and then learn to predict the masked information jointly using the other modalities. We first match the information of query conditions, join conditions, table names, column names, and operations between the query statement and execution plan. Additionally, we match specific numerical information between the execution log and the execution plan. Then, we randomly select some matched parts as the critical information and mask the information in one of the modalities. Specifically with a 50\% probability, we choose to mask tokens in either the query statement or the execution plan. The corresponding tokens will be masked and serve as reference prediction information. 
For the execution logs, we mask parts of the logs and utilize information from query statements and execution plans to predict the masked parts of the logs. This approach enables the model to align query statements, execution plans, and execution logs at the encoding level. For example, we replace the table name "user" in the query statement with "<mask>" while keeping the "user" unchanged in the execution plan. By using the masked query statement, execution plan, and execution logs, we predict the masked token. Following this, we employ a transformer module $A$ to aggregate the three modalities and jointly predict the masked content. Take the \verb|SQL| modality as an example and let $\tilde{S}$ denote its masked content and $S-\tilde{S}$ denote its remaining content. The joint prediction objective for self-supervised pre-training is applied to each modality to achieve fine-grained cross-modal alignment:
\begin{equation}
\label{eq:pretrain_loss_SPL}
\begin{aligned}
    \mathcal{L}_{\text{SQL}} &=\left\|\tilde{S} - A\left(\text{Enc}_S(S-\tilde{S}), \text{Enc}_P(P), \text{Enc}_L(L)\right)\right\|_{2}^{2}\\
    \mathcal{L}_{\text{PLAN}} &=\left\|\tilde{P} - A\left(\text{Enc}_S(S), \text{Enc}_P(P-\tilde{P}), \text{Enc}_L(L)\right)\right\|_{2}^{2}\\
    \mathcal{L}_{\text{LOG}} &=\left\|\tilde{L} - A\left(\text{Enc}_S(S), \text{Enc}_P(P), \text{Enc}_L(L-\tilde{L})\right)\right\|_{2}^{2}
\end{aligned}
\end{equation}

The \verb|KPIs| modality contains information about the current database state. Considering the impact of the database state on query execution, it is crucial to capture both temporal features of the database state and the patterns of interaction between different metrics of the database instance. 
KPIs represent the database state, and we explore their temporal characteristics. To capture patterns in metrics, we pre-train the embeddings of metrics separately.
Inspired by this property, we propose to employ an auto-encoding structure and reconstruction of the KPI time series as the pre-training objective:
\begin{equation}
\label{eq:pretrain_loss_I}
    \mathcal{L}_{\text{KPIs}} = \left\|\text{Dec}_{I}\left(\text{Enc}_{I}(I)\right) -I\right\|_{2}^{2},
\end{equation}
where $\text{Dec}_{I}$ is the decoder that is to reconstruct the original KPI input from its embeddings. Through this pre-training process, we can obtain better embeddings for KPIs related to anomaly detection, which are very relevant to the downstream task of slow queries, i.e., estimating the impact of root causes.

\textit{Pre-training with large quantities of queries:} As the pretraining of the encoders is self-supervised and only uses the queries themselves, not their root cause impact, we can use all queries, not only the slow queries, in the root cause impact dataset $D$. Thus, we use a large quantity of queries collected from the cloud database system to facilitate the pre-training.

\begin{figure}[t]
\begin{center}
  \includegraphics[width=0.8\linewidth]{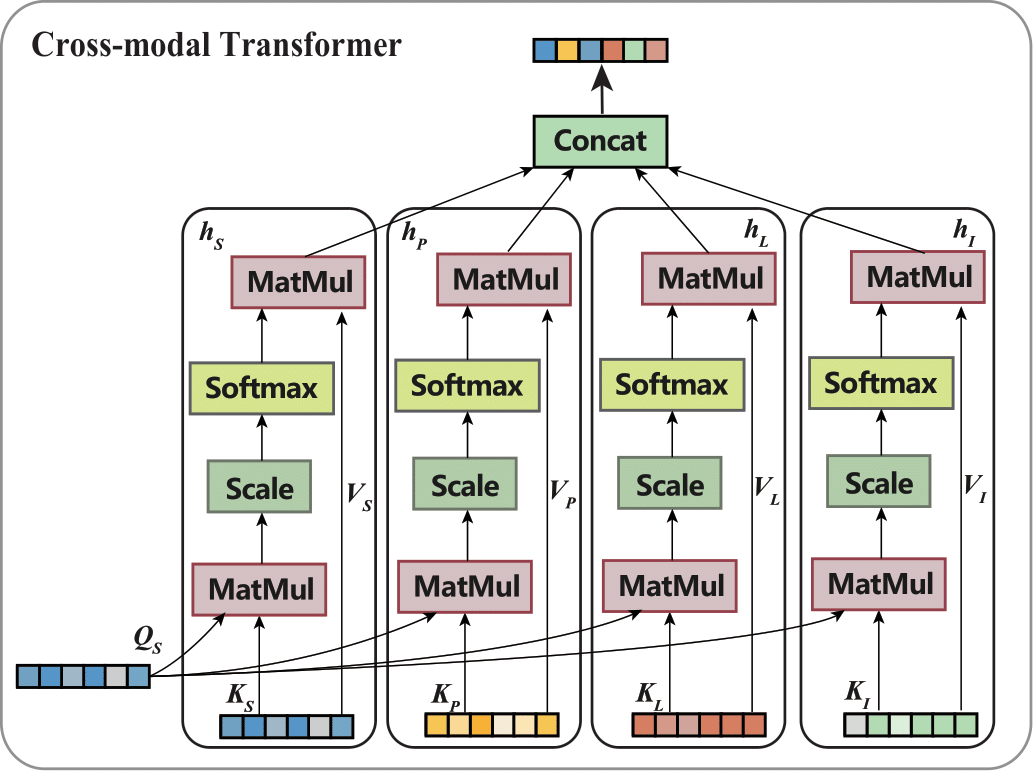}
\end{center}
  \caption{Architecture of the cross-modal Transformer.} 
\label{fig:CrossTransformer}
\end{figure}

\subsubsection{Root-Cause-Adaptive Cross Transformer}\label{sec:multimodal fusion}
\ 
\newline \textbf{Cross-modal Transformer.} To more fully utilize complementary information from the diverse modalities, the multimodal fusion module learns to fuse them to extract features for root cause diagnosis. We employ the Transformer~\cite{DBLP:conf/nips/transformer} architecture and cross-modal attention to integrate information from multiple modalities.

Cross-attention ($CA$)~\cite{DBLP:conf/nips/ALBEF} uses the same computational method as self-attention to process the relationship between different input data A and B:
\begin{equation}
\begin{aligned}
    CA(A, B)=\text{softmax}\left(\frac{AW_{A}^{Q}BW_{B}^{K}}{\sqrt{d_q}}\right)BW_{B}^{V},
\end{aligned}
\end{equation}
where $W_{A}^{Q}$ is the query projection layer of A, $W_{B}^{K}$ is the key projection matrix of B, and 
$W_{B}^{V}$ is the value projection matrix of B.

Computing cross-attention for each pair of modalities incurs large computational costs and introduces redundancy in the fused features. As shown in Figure~\ref{fig:CrossTransformer}, we thus instead design the cross-modal Transformer to have the query statement as the main modality and to serve as the query in cross-attention, while other modalities serve as the keys and values in attention. The choice of the main modality is based on the insight that the \verb|SQL| modality is directly related to a diagnosed slow query. Later, we report on experiments that validate this design.

Take the cross-modal attention between the query statement $S$ and the execution plan $P$ as an example and let $E_S$ and $E_P$ denote the embeddings of the two modalities. We capture cross-modal dependencies between these two and then extract the cross-modal feature $h_P$ from the plan modality:
\begin{equation}
\begin{aligned}
Q_S=E_SW_S^{Q},K_P=E_PW_P^{K},V_P=E_PW_P^{V},
h_P = \text{softmax}(\frac{Q_SK_P^T}{\sqrt{d}})V_P,
\end{aligned}
\end{equation}
where $W^Q_{S}, W^K_{P}, W^V_{P} \in \mathbb{R}^{d \times d}$ are learnable feature projection and $d$ is the dimension of the input embeddings. Applying similar calculations, we can obtain the cross-attention features from all the modalities: $h_S$, $h_P$, $h_L$, $h_{I}$. Having extracted the multimodal features that capture cross-modal relationships, the cross-modal Transformer ($\text{CMT}$) finally concatenates these along the feature dimension and obtains the fused representation through a feed-forward network ($\text{FFN}$):
\begin{equation}
\begin{aligned}
    \text{CMT}(E_S, E_P, E_L, E_I) = \text{FFN}([h_S; h_P; h_L; h_{I}]) 
\end{aligned}
\end{equation}

\textbf{Root-Cause-Adaptive Fusion.} Features from multimodal fusion give a more comprehensive view of the slow queries, which helps the more accurate diagnosis of their root causes. In practice, as the query becomes more complex (e.g., with longer query statements, more nodes in the execution plan, larger numbers of returned rows, and longer execution time), only some parts of the information are highly related to the slow query and should be focused on in diagnosis. Furthermore, different root causes may have diverse relationships with different parts of multiple modalities.
For example, when the root cause is a bad index, more attention should be given to the number of rows in the indexed field and the indexing method. This inspires us to fuse multimodal embeddings adaptively for diagnosing distinct root causes, as illustrated in Figure~\ref{fig:Overview}.

To distinguish the commonality and specificity of multimodal fusion among root causes, we decompose the features into two parts. One part involves features related to the commonality of the root causes, which are shared by all root causes. The other part involves features related to the specificity of each root cause, which are determined by their unique characteristics. We employ the Common Cross-Modal Transformer (denoted as $\text{C-CMT}$) to extract the fused representation $E_F^{C}$ of the common features:
\begin{equation}
\label{eq:C_CMT}
  E^{C}_{F} = \text{C-CMT} (E_S, E_P, E_L, E_{I})
\end{equation}
To extract specific features for each root cause, we employ an Adaptive Cross-Modal Transformer (denoted as $\text{A-CMT}$) that adaptively fuses multimodal embeddings based on distinct root causes. For each root cause $RC_j$, we propose a gating mechanism to control the embedding of each modality during fusion. Take the query statement $E_S$ as an example: the gating unit $G_S^j$ outputs the gating value as the weight and then selects more relevant information for fusion:
\begin{equation}
\label{eq:gate}
  G_S^j(E_S) = \text{sigmoid}\left(\text{FC}(E_S)\right) \odot E_S,
\end{equation}
where $\text{sigmoid}$ is the activation function, $\odot$ is the element-wise product, and $\text{FC}$ is the submodule that output the gating value; we use a linear neural network layer for simplicity. Embeddings controlled by the gating unit then go through the cross-modal Transformer to extract specific multimodal features of the root cause $RC_j$:
\begin{equation}
\label{eq:A_CMT}
  E_A^{j} = \text{A-CMT}\left(G_S^j(E_S), G_P^j(E_P), G_L^j(E_L) G_I^j(E_I)\right).
\end{equation}
The final adaptive-fusion feature $E_F^{j}$ for estimating the root cause $RC_j$ is computed as the sum of the common feature $E_{C}$ and the root-cause-adaptive feature $E_{A}^{j}$:
\begin{equation}
\label{eq:csfusion}
  E_F^{j} = E_{C} + E_{A}^{j}
\end{equation}

\subsubsection{Unified Identification and Ranking}\label{sec:root cause prediction}
\ 
\newline \textbf{Estimation of root cause impacts.} In order to identify and rank valid root causes according to their influence on the slow query in a unified framework, we propose a training objective for estimating the impact of each potential root cause. 

Given the multimodal input data $X_i$, the input embedding module and the multimodal fusion module extract the multimodal feature $E_F=\left\{E_F^{j}\right\}_{j=1}^{r}$ as detailed in Sections~\ref{sec:input embedding} and~\ref{sec:multimodal fusion}, where $E_F^{j}$ is the adaptively extracted feature for the root cause $RC_j$ and $r$ is the total number of all potential root causes. With $E_F^{j}$ as input, the root cause estimation module estimates the impact of each root cause $RC_j$ as $\hat{y}_{ij}$:
\begin{equation}
    \hat{y}_{ij} = \text{MLP} (E_F^{j})
\end{equation}
During training, we minimize the mean squared error (MSE) between the estimated and the annotated impacts of all the potential root causes:
\begin{equation}
    \mathcal{L}_\text{pred} = \sum_{j=1}^{r} \Vert y_{ij} - \hat{y}_{ij} \Vert_2^2
\end{equation}
Recall that we regard root causes as valid if their impact exceeds a threshold $\epsilon$. Here, we use the same threshold $\epsilon$ to apply the estimated impacts to obtain valid root causes and then rank them based on the estimated impacts. 

\textbf{Impact-aware Regularization.} To further improve the accuracy of root cause identification and ranking, we propose two impact-aware regularizations to combine with the MSE training loss to obtain the overall training loss of the framework. 
The identification of root causes may be influenced by noise in the estimated impact, especially noise near the threshold $\epsilon$. Motivated by this, we propose the valid regularization: 
\begin{equation}
\begin{aligned}
    \mathcal{L}_\text{valid} = \sum_{j=1}^{r} \max(0, \mathbb{I}(y_{ij} < \epsilon) \cdot (\hat{y}_{ij} - \epsilon) + \eta ), 
\end{aligned}
\end{equation}
where $\eta$ is a margin, $\mathbb{I}(y_{ij} < \epsilon)$ is an indicator function that returns $1$ if $y_{ij} < \epsilon$ and $-1$, otherwise. This encourages the estimated impact to be larger or smaller than threshold $\epsilon$ by at least a margin of $\eta$, depending on whether or not the root cause is valid. This regularization enhances the ability to distinguish between valid and invalid impacts in root cause identification.

Optimizing the model with only the MSE loss may not ensure an accurate ranking of the root causes due to the imperfect fit and estimations. We propose an order regularization to constrain the rankings of root causes. With $\{z_{ij}\}_{j=1}^{r}$ denoting sorted root causes according to ground truth impacts $y_{ij}$ and $\{\hat{z}_{ij}\}_{j=1}^{r}$ denoting sorted root causes according to estimated impacts $\hat{y}_{ij}$, the regularization is:
\begin{equation}
    \mathcal{L}_\text{order} = \sum_{j=1}^{r-1}{\max(0, \left((z_{ij} - z_{i(j+1)}) - (\hat{z}_{ij}-\hat{z}_{i(j+1)})\right))}
\end{equation}
The regularization constrains estimated orders to at least follow the margins between the ground truth orders, thereby preserving the impact ranking of the root causes. For example, given two root causes that have adjacent rankings, if the distance $\hat{z}_{ij}-\hat{z}_{i(j+1)}$ between the estimated results is 6.5\% and the ground truth distance $z_{ij} - z_{i(j+1)}$ is 9.6\%, we must update the estimated impact to be closer to the ground truth distance. This is so because the estimated distance is less than the ground truth distance, which affects the estimation of important root causes and even changes the order of root causes. Conversely, if the distance $\hat{z}_{ij}-\hat{z}_{i(j+1)}$ between the estimated results is 10.1, which exceeds the ground truth distance $z_{ij} - z_{i(j+1)}$ of 9.6, we does not perform any adjustment as this does not lead to incorrect estimation of important root causes or their order. The overall training loss of the framework is formulated as follows.
\begin{equation}
\label{eq:train_loss}
    \mathcal{L} = \mathcal{L}_\text{pred} + \lambda(\mathcal{L}_\text{valid} + \mathcal{L}_\text{order}),
\end{equation}
where $\lambda$ is a trade-off hyper-parameter between the estimation loss and the impact-aware regularization losses. 

\subsubsection{Inference}
We design an estimation model to estimate root cause impacts and rank them to determine the key root causes. As shown in Algorithm~\ref{algo:MD4RC}, given a slow query $X_i$, they are processed individually through the pre-trained models $\text{ENC}_{S}$, $\text{ENC}_{P}$, $\text{ENC}_{L}$, and $\text{ENC}_{I}$ to obtain $E_{S}$, $E_{P}$, $E_{L}$, and $E_{I}$. Using $\text{C-CMT}$, we extract the common features shared by each root cause. Then, we utilize a gating mechanism to capture the different parts of the query statement, execution plan, execution log, and KPIs that each root cause focuses on. For each root cause, we use $\text{A-CMT}$ to extract its specific features. We combine the common and specific features of each root cause and employ a Linear layer to estimate the impact of each individual root cause.  Finally, we sort all root causes by their impact $y_{ij}$ and eliminate those that are below impact threshold $\epsilon$ to obtain a list of valid root causes and their impact.

\IncMargin{1em}
\begin{algorithm}  \SetKwInOut{Input}{input}\SetKwInOut{Output}{output} 

\SetKwFunction{FuseFeatures}{FuseFeatures} 
\SetKwProg{Fn}{Function}{:}{\KwRet} 
	 
      \textbf{Pre-training:} \\
      \textbf{Input:} All queries used for pre-training $\{Q_i|i \in [N] \}$ \\
      Initialize the encoder $\mathbb{E}_{\phi}$\;
      $\tilde{S},\tilde{P},\tilde{L}$ $\leftarrow$ Mask $S,P,L$ in $Q_i$\;
	 \For{$i\leftarrow 1$ \KwTo $N$}{
      \For{$\tilde{T} \in \{\tilde{S}, \tilde{P}, \tilde{L}\}$}{
        Encode $\tilde{T}$ and other original inputs using $\mathbb{E}_{\phi}$\; 
      }
      Reconstruct $I$ in $Q_i$ using encoder $\mathbb{E}_{\phi}$\;
      }
      Update ${\phi}$ based on Eqs.~\ref{eq:pretrain_loss_SPL} and ~\ref{eq:pretrain_loss_I}\;
          
 \BlankLine
  \BlankLine
\textbf{Training}: \\
      \textbf{Input:} A root cause impact dataset $D=\{(y_{ij}, X_i, RC_j)|i \in [M], j \in [r]\}$ \\
      Initialize the weights $\theta$ of the whole estimation model with its encoder's weights ${\phi}$ initialized as the pre-trained weights\;
      \For{$i\leftarrow 1$ \KwTo $M$}{
      $E_S, E_P, E_L, E_{I} \leftarrow$ Encode $S,P,L,I$ in $X_i$\;
      $E_{F}^{C}$ $\leftarrow$ Extract common features based on Eq.~\ref{eq:C_CMT}\;
      
      \For{$j\leftarrow 1$ \KwTo $r$}{ 
      \FuseFeatures{$E_S, E_P, E_L, E_{I}, E_{F}^{C}$}\;
      $\hat{y}_{ij} \leftarrow$ Estimate impact of root cause $RC_{j}$\;
      }
      }
    Update $\theta \leftarrow \arg\min_\theta(\mathcal{L}(y_{ij}, \hat{y}_{ij}))$ using Eq.~\ref{eq:train_loss}\;
  
  \BlankLine
  \BlankLine
  
        \textbf{Inference}: \\
      \textbf{Input:} A new slow query $X_i$  \\

        $E_S, E_P, E_L, E_{I} \leftarrow$ Encode $S,P,L,I$ in $X_i$\;
        $E_{F}^{C}$ $\leftarrow$ Extract common features based on Eq.~\ref{eq:C_CMT}\;
        \For{$j\leftarrow 1$ \KwTo $r$}{ 
        \FuseFeatures{$E_S, E_P, E_L, E_{I}, E_{F}^{C}$}\;
        $\hat{y}_{ij} \leftarrow$ Estimate impact of root cause $RC_{j}$\;
        }
	 
      \textbf{Return:} Validate the root cause using $\hat{y}_{ij}$ and sort $\hat{y}_{ij}$ in descending order
      
  \BlankLine
  \BlankLine

\Fn{\FuseFeatures{$E_S, E_P, E_L, E_{I}, E_{F}^{C}$}}{

    Calculate the gate value using Eq.~\ref{eq:gate}\;
    Extract specific features based on Eq.~\ref{eq:A_CMT}\;
    \textbf{Return:} {Fuse the common and specific features using Eq.~\ref{eq:csfusion}}\;
            
}
      
 	 	  \caption{Pseudo-code of \texttt{RCRank}}
      
 	 	  \label{algo:MD4RC} 
 	 \end{algorithm}
 \DecMargin{1em}

\section{Experiments}
\subsection{Experimental Design}
\subsubsection{Datasets}

We collect five datasets, Hologres1, Hologres2, TPC-DS, TPC-C, and TPC-H, with the slow query threshold $\delta$ set to 1 second and the valid root cause threshold $\epsilon$ set to 10\%. 
We consider 5 types of root causes in Hologres1, Hologres2, and TPC-DS: 1. outdated statistical information; 2. under-optimized join order algorithm; 3. missing or redundant indexes; 4. inappropriate distribution keys; 5. poorly written queries.  
In TPC-C and TPC-H, we consider 10 types of root causes: 1. outdated statistical information; 2. under-optimized join order; 3. inappropriate distribution keys; 4. missing indexes; 5. redundant indexes; 6. repeatedly executing subqueries; 7. complex table joins; 8. updating an entire table;  
9. inserting large data; 10. unknown root causes.
We provide detailed statistics of the datasets in Table~\ref{table:Dataset}, including the total number of queries (\#Queries) that we monitored, the number of slow queries (\#Slow Queries), and the number of tables (\#Tables).

(1) \textbf{Hologres 1 \& 2}: We collect two datasets of slow queries from Alibaba's production cloud database system Hologres, which contains a large number of real-world queries. The root cause distribution in dataset \textbf{Hologres1} tends to lean towards lacking indexes, while the root cause distribution in \textbf{Hologres2} is uniform. The slow queries come from routine executions of Alibaba's internal activities, involving databases with millions of records and complex conditions. We employ the rule-based and LLM-based methods outlined in Section~\ref{sec:collection} to collect the impacts of root causes.

(2) \textbf{TPC-DS, TPC-C, and TPC-H}: 
These are standard benchmarks for evaluating the performance of database management systems. TPC-DS simulates a retail enterprise's decision support environment, encompassing complex queries, data mining, and business intelligence functionalities. TPC-C simulates the activities of a wholesale supplier. TPC-H is a decision support benchmark designed for the retail industry. It is possible to generate databases, tables, and specific mock data of chosen sizes. Although TPC-DS, TPC-C, and TPC-H offer query generation, only about a hundred generation templates is insufficient for our experimental needs. We thus use LLMs to generate more complex queries based on the relevant table structures in Postgres. Specifically, we use table structures such as table names and column names as prompts and require the LLM to generate complex query statements. After a selection process, we use the revision methods from Section~\ref{sec:collection} to perform root cause annotation on the slow queries, thereby producing synthetic datasets.

\begin{table}[t]
\centering
\caption{Details of datasets.}
\resizebox{0.40\textwidth}{!}
{
\begin{tabular}{c|c|c|c} 
\toprule
\textbf{Dataset} &  \textbf{\#Queries}  & \textbf{\#Slow Queries}   & \textbf{\#Tables}  \\ 
\hline
Hologres1           &     12631    & 2209                          &     136           \\
\hline
Hologres2      &      13544        &     2701                          &     84           \\
\hline
TPC-DS        &      12196      & 7246                              & 24                    \\ 
\hline
TPC-C      &      9980        &     5764                          &     10           \\
\hline
TPC-H      &      7638        &     4279                          &     8           \\
\bottomrule
\end{tabular}
\label{table:Dataset}
}
\end{table}

\subsubsection{Baseline Approaches}
(1) \textbf{OpenGauss}~\cite{DBLP:journals/pvldb/Opengauss}: The SQL-level Diagnosis component in OpenGauss is similar to our task, which uses  
execution plans, and key performance indicators as input to find a time-consuming SQL operator. However, OpenGauss cannot rank the root causes. To enable OpenGauss to estimate an impact of the root cause, we fuse its execution plan and KPI features and use a fully connected layer for impact estimation.
(2) \textbf{D-BOT}~\cite{DBLP:journals/corr/DBot}: D-BOT collects diagnostic files and utilizes an LLM to automatically extract knowledge from these to diagnose the root causes of slow queries.
(3) \textbf{Only-SQL, Only-Plan,
Only-KPI, Only-Log}: We use the embedding module of our proposal to learn representations of a single modality and then estimate the impact of the root cause through a fully connected layer.
(4) \textbf{Concat}: We use the embedding modules of our proposal to learn the representation of each modality and then concatenate them and estimate the impact of the root cause through a fully connected layer.

\subsubsection{Evaluation Metrics}
We evaluate the effectiveness of the methods based according to four metrics: impact error, root cause accuracy, root cause order, and key root cause.

First, we evaluate whether the estimated root cause impact values are accurate compared to ground truth impact values. We compute MSE error for each estimated impact values and then report the mean and standard deviation of the estimated impact values in the whole test set.

Second, we evaluate the accuracy of the valid root causes ($\textit{V-ACC}$) and the most important valid root causes ($\textit{Top1-ACC}$). Recall that a root cause is valid if its impact value exceeds the root cause threshold $\epsilon$. Specifically, \textit{V-ACC} calculates the accuracy of valid root causes by computing the ratio of correctly estimated root causes to the total number of root causes.
\textit{Top1-ACC} calculates the accuracy of the root cause with the highest impact value.

Third, we evaluate the root cause orders obtained by the estimated vs. the ground truth impacts. We employ the multi-cause accuracy ($\textit{MC-ACC}$) and Kendall's Tau (\textit{Tau}) metrics. \textit{MC-ACC} measures the accuracy of estimating the list of valid root causes sorted by the impact of each root cause for individual queries. Specifically, $\textit{MC-ACC}=\frac{\text{Order}_{RC}}{m}$, where $\text{Order}_{RC}$ is the number of lists of estimated valid root causes that are ordered consistently with the ground truth. 
\textit{Tau} evaluates the performance in terms of order accuracy: 
$Tau(RC) = \frac{n_c-n_d}{\sqrt{(n_0-n_a)(n_0-n_b)}}$, where $n_c$ is the number of concordant pairs where the estimated impact values are consistent with the actual values in their order; $n_d$ is the number of discordant pairs. 
$n_a$ and $n_b$ represent the number of tied rankings in the ordered lists of estimated root causes and ground truth root causes, respectively; and $n_0 = \frac{n(n-1)}{2}$, where n is the length of the list. In practical applications, the negative impact of providing incorrect root causes exceeds the negative impact of under-estimated root causes.

Finally, to measure the improvement the estimated top-1 root cause can bring, we use $\textit{Top1-IR}$ to evaluate the performance gain after revising queries according to the top-1 root cause. Specifically, $\textit{Top1-IR}$ calculates the impact value of the estimated top-1 root cause for each slow query on average. In the process of revising slow queries, users prioritize revising slow queries according to the root cause with the greatest impact, i.e., the top-1 root cause. Higher $\textit{Top1-IR}$ values indicate that the top-1 root cause yields a larger performance improvement. Thus, the higher the $\textit{Top1-IR}$, the better. 

In Tables~\ref{table:mainResult}, ~\ref{table:mainResulttpctph}, ~\ref{table:mainResultSynthestictpcds}, ~\ref{table:ablationStudyResult}, and ~\ref{table:mainModalStudyResult}, $\uparrow$ indicates that higher values represent better performance, while $\downarrow$ indicates that lower values represent better performance.

\subsubsection{Implementation}
We split the root cause impact dataset $D$ into 8:1:1 for training, validation, and testing. We use all queries, except the validation and testing part of the slow queries, for pre-training encoders. 
We conduct training on an NVIDIA GeForce RTX 3090 GPU with a batch size of $64$ and train for $50$ epochs. We employ bert-base-uncased~\cite{DBLP:conf/naacl/Bert} for encoding queries, Queryformer~\cite{DBLP:journals/pvldb/Queryformer} for encoding plans, a multi-layer perceptron for encoding logs, and a 2DCNN for encoding KPIs. The bert-base-uncased uses pre-trained parameters and the multi-layer perceptron consists of three fully connected layers with 13, 64, and 32 dimensions for the three layers. In the Cross Transformer, we set the number of blocks to $3$ and the dropout rate to $0.1$. We utilize the Adam optimizer with $\beta_1=0.9$, $\beta_2=0.999$, and a learning rate of $lr=0.0003$. The trade-off for the losses is set to $\lambda=7$, with further considerations of the value of $\lambda$ in Section~\ref{sec:sensitivity}.

\begin{table*}[t]
\centering
\caption{Overall root cause diagnosis results in Hologres.}
\resizebox{0.98\textwidth}{!}
{
\begin{tabular}{c|cccc | cccccc | cccccc} 
\toprule
  & \multicolumn{4}{c|}{Modality} & \multicolumn{6}{c|}{Hologres1} & \multicolumn{6}{c}{Hologres2} \\
\hline
Method        
& SQL      & Plan         & Log          & KPIs
& V-ACC $\uparrow$           & Top1-ACC $\uparrow$           & MSE$\pm$std $\downarrow$    & MC-ACC $\uparrow$      & Tau $\uparrow$   & Top1-IR $\uparrow$    & V-ACC $\uparrow$           & Top1-ACC $\uparrow$           & MSE$\pm$std $\downarrow$      & MC-ACC $\uparrow$      & Tau $\uparrow$   & Top1-IR $\uparrow$     \\
\hline
Concat      &    \Checkmark    & \Checkmark &    \Checkmark  & \Checkmark
                                       & \underline{0.7482}          & \underline{0.5071}          & \underline{0.1847$\pm$0.1813}        & 0.2132          & \underline{0.3074}          & 0.1751        & \underline{0.7254} & \underline{0.4240} & 0.2274$\pm$0.1793   & \underline{0.3253} & \underline{0.3319}  & \underline{0.2364} \\
Only-SQL              &    \Checkmark    &  &     &
     & 0.6898          & 0.4067          & 0.3595$\pm$0.1953  &      0.1955          & 0.2269          & 0.1632    & 0.6989 & 0.3696 & \underline{0.1056$\pm$0.1954} & 
0.2659 & 0.1748 &  0.1508 \\
Only\_Plan     &      & \Checkmark &      &
    & 0.7340          & 0.4743          & 0.2236$\pm$0.1899      & \underline{0.2281}          & 0.2429          & 0.1680     & 0.6973 & 0.3803 & 0.4166$\pm0.1987$ & 0.2925 & 0.2059 & 0.1584   \\
Only-Log      &       &  &    \Checkmark  &                                  & 0.6666          & 0.2900          & 1.8796$\pm$1.2394       & 0.1613          & 0.1636          & 0.1592   & 0.6883 & 0.3590 & 0.1747$\pm$0.2045 & 0.2659 & 0.2165 &  0.1422 \\
Only-KPI   &        &           &    & \Checkmark                                  & 0.7058          & 0.3921
 & 0.3989$\pm$0.1854   &   0.1773          & 0.2230          & 0.1557       & 0.6922 & 0.3723 & 0.1522$\pm$0.1923 & 0.2686 & 0.2093 & 0.1410 \\
D-BOT (Vicuna)           &    \Checkmark    & \Checkmark &     \Checkmark   & \Checkmark
                                   & 0.3769          &      -           &    -            &   -           &   -              &    -      & 0.3518 & - & - & - & - & -  \\
OpenGauss   &        & \Checkmark &      & \Checkmark
               & 0.7354          & 0.5032          & 0.4362$\pm$0.1870        & 0.2019          & 0.2740          & \underline{0.1789}   & 0.7149 & 0.4069 & 0.2335$\pm$0.1847 & 0.3058 & 0.2518 & 0.1919  \\
\hline

\begin{tabular}[c]{@{}l@{}}\texttt{RCRank} \end{tabular} &    \Checkmark    & \Checkmark &    \Checkmark  & \Checkmark & \textbf{0.7628} & \textbf{0.5384} & \textbf{0.1226$\pm$0.1752} & \textbf{0.2371} & \textbf{0.3518} & \textbf{0.1988}  & \textbf{0.7420} & \textbf{0.4335} & \textbf{0.0896$\pm$0.1645} & \textbf{0.3563} &
\textbf{0.3611} & \textbf{0.2764} \\

\hline
Improvement &        &  &      &   & \textbf{1.95\%} & \textbf{6.17\%} & \textbf{33.62\%} & \textbf{3.95\%} & \textbf{14.44\%} & \textbf{11.12\%}  & \textbf{2.29\%} & \textbf{2.24\%} & \textbf{15.15\%} & \textbf{9.53\%} &
\textbf{8.80\%} & \textbf{16.92\%} \\
\bottomrule
\end{tabular}
\label{table:mainResult}
}
\end{table*}

% TPC-C、TPC-H
\begin{table*}[t]
\centering
\caption{Overall root cause diagnosis results in TPC-C and TPC-H.} 
\resizebox{0.98\textwidth}{!}
{
\begin{tabular}{c| cccccc | cccccc} 
\toprule
 & \multicolumn{6}{c|}{TPC-C} & \multicolumn{6}{c}{TPC-H}\\
\hline
Method        
& V-ACC $\uparrow$           & Top1-ACC $\uparrow$           & MSE$\pm$std $\downarrow$         & MC-ACC $\uparrow$      & Tau $\uparrow$   & Top1-IR $\uparrow$   & V-ACC $\uparrow$           & Top1-ACC $\uparrow$           & MSE$\pm$std $\downarrow$         & MC-ACC $\uparrow$      & Tau $\uparrow$  & Top1-IR $\uparrow$ \\
\hline

Concat&\underline{0.8340}&\underline{0.5619}&\underline{0.1814$\pm$0.1438}&\underline{0.4917}&0.4722&\underline{0.1413}&\underline{0.8277}&\underline{0.5743}&\underline{0.1943$\pm$0.1855}&0.4253&\underline{0.4944}&\underline{0.1677}\\
Only-SQL&0.7602&0.5281&0.2794$\pm$0.1774&0.4464&0.3965&0.1203&0.7425&0.5399&0.2914$\pm$0.2216&0.4587&0.3492&0.1375\\
Only\_Plan&0.7808&0.5307&0.2681$\pm$0.2349&0.4791&0.4177&0.1294&0.7714&0.5344&0.2731$\pm$0.1841&\underline{0.4757}&0.4430&0.1294\\
Only-Log&0.7422&0.4711&0.3105$\pm$0.2847&0.4359&0.4207&0.1176&0.7235&0.4933&0.3142$\pm$0.2763&0.4188&0.3241&0.1025\\
Only-KPI&0.8054&0.5402&0.2347$\pm$0.1547&0.4224&0.4796&0.1227&0.7827&0.5277&0.2467$\pm$0.2164&0.4579&0.4428&0.1103\\
D-BOT(GPT-3.5)&0.6503&-&-&-&-&-&0.6108&-&-&-&-&-\\
OpenGauss&0.8216&0.5416&0.2103$\pm$0.1697&0.4715&\underline{0.5176}&0.1374&0.8146&0.5633&0.2120$\pm$0.1977&0.4187&0.4703&0.1522\\
\hline
\begin{tabular}[c]{@{}l@{}}\texttt{RCRank}\end{tabular}&\textbf{0.8611}&\textbf{0.6417}&\textbf{0.1221}$\pm$0.1378&\textbf{0.5247}&\textbf{0.5574}&\textbf{0.1564}&\textbf{0.8523}&\textbf{0.6512}&\textbf{0.1496}$\pm$0.1672&\textbf{0.5718}&\textbf{0.5125}&\textbf{0.1712}\\
\hline
Improvement &  \textbf{3.25\%} & \textbf{14.20\%} & \textbf{32.69\%} & \textbf{6.71\%} & \textbf{7.69\%} & \textbf{10.69\%} &  \textbf{2.97\%} & \textbf{13.39\%} & \textbf{23.01\%} & \textbf{20.20\%} & \textbf{3.66\%} & \textbf{2.09\%} \\

\bottomrule
\end{tabular}
}
\label{table:mainResulttpctph}

\end{table*}

% TPC-DS
\begin{table}[t]
\centering
\caption{Overall root cause diagnosis results in TPC-DS.} 
\resizebox{0.48\textwidth}{!}
{
\begin{tabular}{c| cccccc} 
\toprule
  & \multicolumn{6}{c}{TPC-DS} \\
\hline
Method        
& V-ACC $\uparrow$           & Top1-ACC $\uparrow$           & MSE$\pm$std $\downarrow$         & MC-ACC $\uparrow$      & Tau $\uparrow$   & Top1-IR $\uparrow$  \\
\hline

Concat&\underline{0.7836}&\underline{0.5566}&0.7133$\pm$0.1732&\underline{0.4547}&\underline{0.4873}&\underline{0.1201}\\
Only-SQL&0.7381&0.5241&0.6133$\pm$0.2677&0.4216&0.3224&0.1005\\
Only\_Plan&0.7644&0.5142& 0.4162$\pm$0.2115&0.4331&0.4574&0.1014\\
Only-Log&0.6984&0.4869&0.9899$\pm$0.4122&0.3824&0.3163&0.0973\\
Only-KPI&0.7713&0.5074& 0.5849$\pm$0.2413&0.4032&0.4130&0.0989\\
D-BOT(GPT-3.5)&0.5813&-&-&-&-&-\\
OpenGauss&0.7683&0.5243& \underline{0.3147$\pm$0.1472}&0.3926&0.4241&0.1142\\
\hline
\begin{tabular}[c]{@{}l@{}}\texttt{RCRank}\end{tabular}&\textbf{0.8466}&\textbf{0.6321}&\textbf{0.1732}$\pm$0.1440&\textbf{0.5431}&\textbf{0.5391}&\textbf{0.1477}\\
\hline
Improvement &  \textbf{8.04\%} & \textbf{13.56\%} & \textbf{44.96\%} & \textbf{19.44\%} & \textbf{10.63\%} & \textbf{22.98\%} \\

\bottomrule
\end{tabular}
}
\label{table:mainResultSynthestictpcds}

\end{table}

\begin{table*}[t]
\centering
\caption{Ablation studies on Hologres and TPC-DS.}
\resizebox{0.95\textwidth}{!}
{
\begin{tabular}{c|c|c|c|c|c|c|c|c|c|c|c|c} 
\toprule
Dataset & \multicolumn{6}{c|}{Hologres} & \multicolumn{6}{c}{TPC-DS} \\
\hline
Method                                                                                                          & V-ACC $\uparrow$           & Top1-ACC $\uparrow$           & MSE $\downarrow$         & MC-ACC $\uparrow$      & Tau $\uparrow$   & Top1-IR $\uparrow$     & V-ACC $\uparrow$           & Top1-ACC $\uparrow$           & MSE $\downarrow$         & MC-ACC $\uparrow$      & Tau $\uparrow$   & Top1-IR $\uparrow$     \\
\hline
\begin{tabular}[c]{@{}c@{}} w/o gate unit\end{tabular}              & 0.7328 & 0.4400   & 0.1876 &  0.2340     & 0.2688        & 0.1954 & 0.8141 & 0.5907   & 0.1864 & 0.4766     & 0.4998        & 0.1351         \\
\begin{tabular}[c]{@{}c@{}}use concat \end{tabular}                &0.7355&     0.4548     &    0.2342    &    0.2688        &          0.3025    &0.1923  &0.8154&     0.5942     &    0.2126    &    0.5074        &          0.5043     &0.1403          \\
\begin{tabular}[c]{@{}c@{}}use MSE loss\end{tabular} &                 0.7238                                                 &   0.4533       &  0.2624      &       0.2441       &       0.2805        & 0.1712  &                 0.8014                                                 &   0.5981       &  0.2867      &     0.4871       &       0.5114        & 0.1389 \\
w/o pre-train                                             & 0.7360 & 0.4746 &  \textbf{0.1119} &  0.2718 & 0.3269 & 0.2264 & 0.8179 & 0.6014 &  0.1931 & 0.5266 & 0.5161 & 0.1412 \\

\begin{tabular}[c]{@{}l@{}}pre-train\end{tabular} & \textbf{0.7465} & \textbf{0.4985} & \underline{0.1164} &  \textbf{0.2732} & \textbf{0.3303} & \textbf{0.2360} & \textbf{0.8466} & \textbf{0.6321} & \textbf{0.1732} &  \textbf{0.5431} & \textbf{0.5391} & \textbf{0.1477}\\
\bottomrule
\end{tabular}
}
\label{table:ablationStudyResult}
\end{table*}

\subsection{Experimental Results}
\subsubsection{Overall Performance}
\label{sec:performance}
Tables~\ref{table:mainResult}, ~\ref{table:mainResulttpctph}, and ~\ref{table:mainResultSynthestictpcds}
show a comparative analysis of the proposed method and the baseline methods across diverse evaluation metrics on two real datasets and a synthetic dataset. Overall, \texttt{RCRank} outperforms baseline methods at root cause identification and ranking, providing evidence that \texttt{RCRank} is effective. Specifically, we observe the following.

\textbf{(1)} The single-modal approach performs relatively well on Only-SQL and Only-Plan, indicating that the estimation of root cause impact is more dependent on \verb|SQL| and \verb|Plan|. However, the single-modal approach lacks other crucial information; thus, these methods do not have a comprehensive understanding of slow queries. The query statement contains table structure information, the execution plan includes details like execution order and operations, the execution log provides information on actual resource consumption during execution, and KPIs reflect the current state of the system. 

\textbf{(2)} OpenGauss, based on the concatenation fusion of execution plan and KPIs, only integrates information from the execution plan and KPIs, lacking details from query statements and useful information from execution logs for identifying root causes. In contrast, our approach utilizes information from query statement, execution plan, execution logs, and KPIs, fusing features through a Cross-modal Transformer across these four modalities. This allows us to make use of additional information relevant to root causes. The table shows that paying attention to information such as the semantic content in query statements and actual execution information in execution logs can enhance the diagnosis of slow query root causes and the estimation of their impact. 

\textbf{(3)} Due to the insensitivity of LLMs to numbers, D-Bot cannot estimate the impact of root causes. Therefore, in D-Bot, we focus on the accuracy of single root cause estimation. Due to privacy concerns, we could not use GPT as the underlying LLM on Hologres. Instead, we opt for the open-source vicuna-13B-1.5k model, which we deploy locally. In scenarios where the use of GPT is not possible, D-Bot exhibits lower accuracy at single root cause estimation. This indicates that for root cause determination, LLMs still require more knowledge and the support of powerful language models as a foundation. On the TPC-DS, we use GPT-3.5 for our large model. We find that the trained model performs better a identifying the root causes of slow queries. 

\textbf{(4)} For the task of identifying and ranking root causes, we focus on the performance improvement of slow queries after applying the revision methods based on root causes. To reduce revision costs, we focus on whether addressing estimating the top-1 root cause can bring substantial improvement. We observe that \texttt{RCRank} outperforms the other methods on the $\textit{Top1-IR}$ metric, providing evidence that integrating query statements, execution plans, execution logs, and KPIs is effective at estimating root causes and impacts. 

\subsubsection{Ablation Studies}
\label{sec:ablation_study}
To study the effect of different components, we create four variants of \texttt{RCRank} and conduct several experiments on both the Hologres and Postgres to compare their performance: 

(1) \textbf{w/o gate unit}: To assess the importance of the Root Cause Gating Unit a providing attention to different parts of the input for different root causes, we remove the gating unit component from the \texttt{RCRank}. As shown in Figure~\ref{table:ablationStudyResult}, we observe that when not using the gate unit, performance is the worst. This suggests that different root causes focus on different parts of the input, e.g., diagnosing different root causes focuses on partial tokens in the query statement, partial operations in the execution plan, partial values in the logs, and partial KPIs in the KPIs. 

(2) \textbf{use concatenation}: To assess the importance of the Cross-modal Transformer at fusing different features, we replace the Cross-modal Transformer with concatenation. This causes a decrease in performance, showing that cross-modal Transformers are more effective at learning integrated features. Cross-fused features can better understand query information, aiding in determining the types and impacts of root causes. 

(3) \textbf{using MSE loss}: To assess the effect of Impact-aware Regularization, we replace the training loss with the MSE loss. We find that $\mathcal{L}_\text{order}$ and $\mathcal{L}_\text{valid}$ play crucial roles in the ordering and accuracy of identifying and ranking the root causes of slow queries. Imposing constraints on the ordering and validity of root causes can enhance the identification of effective and critical root causes. The MSE loss fails to distinguish root causes near the validity threshold. Since the MSE loss only calculates the mean square error, it cannot ensure the correct order of root causes when estimation errors exist.

(4) \textbf{w/o pre-train}: 
To assess the effect of the alignment pretraining method, we compare the performance with and without pre-training. We observe that the pre-training yields an improvement at estimating root causes and their impacts, indicating that using additional queries and aligning query statements, execution plans, and execution logs using the masking method and capturing patterns of KPIs using reconstructing method effectively
capture the features of queries.

\subsubsection{Efficiency.}

\begin{table}
\centering
\caption{Training and inference running times on Hologres.
}
\resizebox{0.35\textwidth}{!}
{
\begin{tabular}{c|c|c} 
\toprule
    & \begin{tabular}[c]{@{}c@{}}
   \textbf{Training time} \\
   (per epoch)
   \end{tabular}   & 
   \begin{tabular}[c]{@{}c@{}}
   \textbf{Inference time} \\
   (per slow query)
   \end{tabular}  \\ 
\hline
\texttt{RCRank}        & 1.76 m                 & 0.018 s                                    \\ 
\hline
OpenGauss         & 1.44 m             & 0.015 s                   \\
\hline
D-Bot       &    -      &     
5.17 m \\
\bottomrule
\end{tabular}
\label{table:Efficiency}
}
\end{table}
Table~\ref{table:Efficiency} shows the training time (per epoch) and inference time (per slow query) of \texttt{RCRank}, OpenGauss, and D-Bot, where we report the average runtime across 5 runs. In the training phase, D-Bot does not require training, since it leverages external knowledge, tools, and the internal knowledge of pre-trained LLMs for diagnosis. Therefore, we only compare the training time of \texttt{RCRank} and OpenGauss. We find that \texttt{RCRank} and OpenGauss have similar training times. 
The additional multimodal inputs do not incur significant computational overhead but result in better estimating the impact of root causes. In the inference phase, the machine learning methods with \texttt{RCRank} and OpenGauss are significantly faster than D-Bot. This is because D-Bot uses its LLM to make multiple calls to its dialogue model for multi-agent joint diagnosis to generate root causes, while \texttt{RCRank} and OpenGauss only require one call to output root causes and their impacts. 

\begin{figure}[t] 
	\centering 
        \subfigure[Hologres]{
		\label{fig:senseHolo}
		\includegraphics[width=0.45\linewidth]{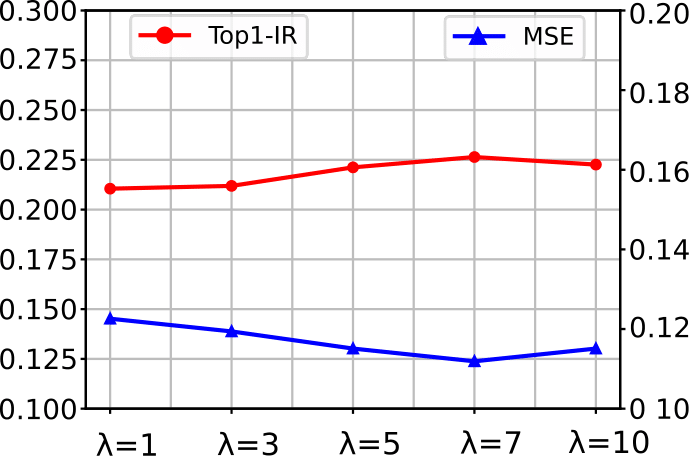}}
	\subfigure[TPC-DS]{
		\label{fig:sensePost}
		\includegraphics[width=0.45\linewidth]{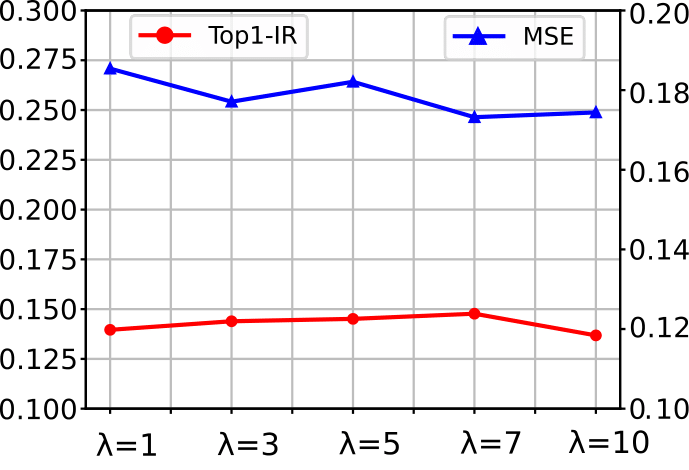}}
	\caption{Sensitivity to parameter $\lambda$ on Hologres and TPC-DS. }
	\label{fig:sensitivity_analysis}
\end{figure}

\subsubsection{Original vs. Improved Run Time of Slow Queries.}
\label{sec:ori_imp}
\begin{figure}[t]
\begin{center}
  \includegraphics[width=0.8\linewidth]{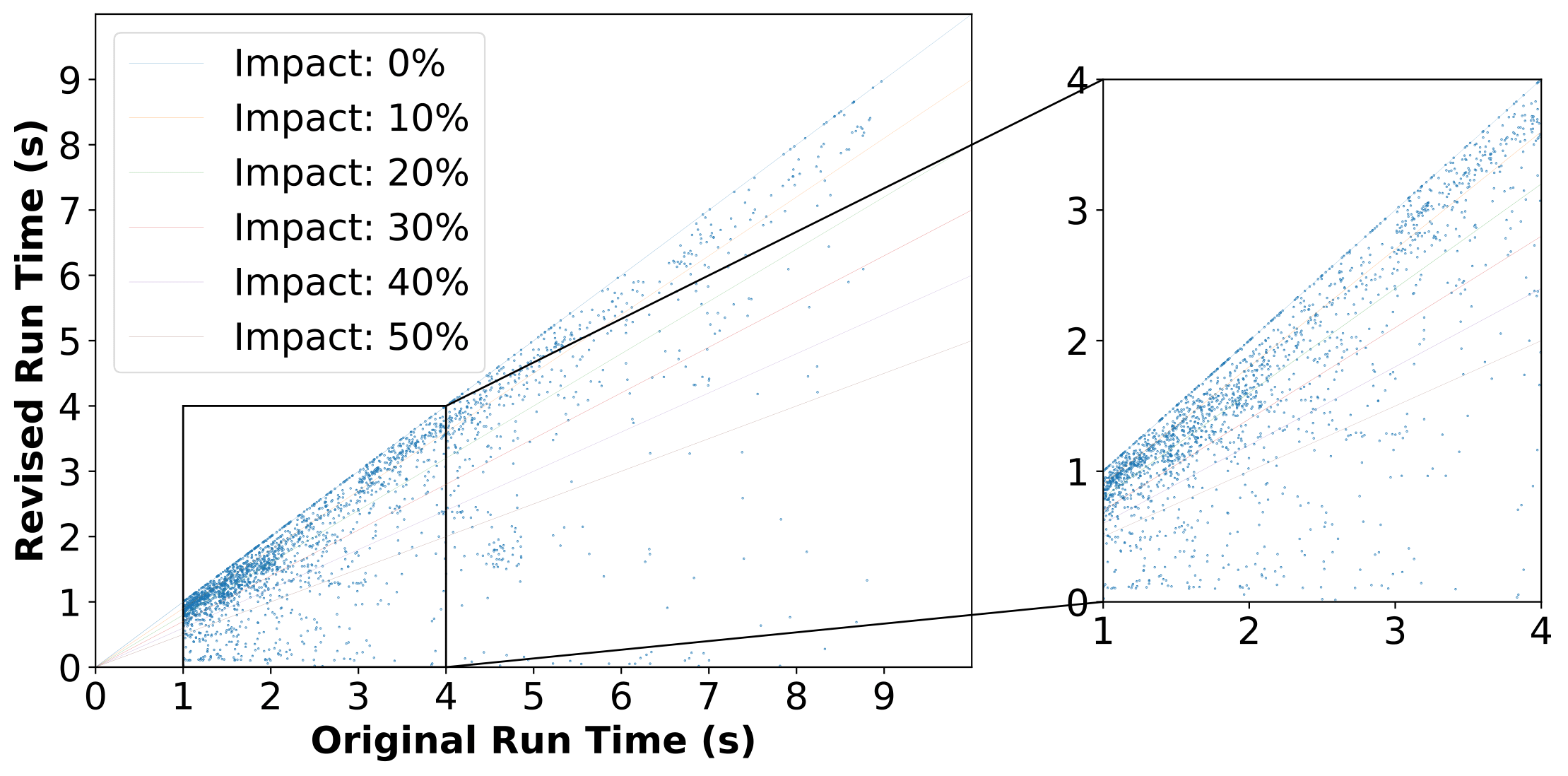}
\end{center}
  \caption{Original vs. improved run time of slow queries in Hologres. 
  }
  
\label{fig:ori_revised_time_paper}
\end{figure}

Our goal is to improve query efficiency by reducing the runtimes of slow queries, rather than completely revising them into non-slow queries. As shown in Figure~\ref{fig:ori_revised_time_paper}, in most cases, the execution times of the revised slow queries are shorter than those of the original slow queries, to varying extents. As shown in Table~\ref{table:run_time_sys_paper}, \texttt{RCRank} identifies root causes that lead to substantial speedups in both Hologres and TPC datasets.

\subsubsection{Sensitivity Analysis}\label{sec:sensitivity}
To study the impact of different $\lambda$ on performance, we investigate the average performance of the Hologres datasets and the performance of the TPC-DS dataset with values of $\lambda$ in $[1,3,5,7,10]$. As shown in Figure~\ref{fig:sensitivity_analysis}, the performance changes only little across different $\lambda$ settings. Because $\mathcal{L}_\text{order}$ constrains both the order of the root causes and the gap between root cause impacts, we obtain more accurate estimates of root cause impacts. 

\begin{table}[t]
\centering

\caption{End-to-end run time improvement.}
\resizebox{0.4\textwidth}{!}
{
\begin{tabular}{c|c|c|c} 
\toprule
Database &    \begin{tabular}[c]{@{}c@{}}Original\\ run time (s)\end{tabular}    & 
   \begin{tabular}[c]{@{}c@{}}Revised\\ run time (s)\end{tabular} & Improvement   \\ 
\hline
Hologres1  & 1736.52                 & 1226.68    &  29.36\%  \\
\hline
Hologres2  & 1297.86                & 901.49   &  30.54\% \\
\hline
TPC-DS  &       1852.54          & 1363.66  &   26.39\% \\
\hline
TPC-C  &      1501.74          &  1092.22 &  27.27\%  \\
\hline
TPC-H  &      1030.33           &  781.202 & 24.18\% \\
\bottomrule
\end{tabular}

\label{table:run_time_sys_paper}
}

\end{table}

\begin{table}[t]
\caption{Main modality studies.}
\resizebox{0.45\textwidth}{!}
{
\begin{tabular}{c|c|c|c|c|c|c} 
\toprule
Modality           & V-ACC $\uparrow$           & Top1-ACC $\uparrow$           & MSE $\downarrow$         & MC-ACC $\uparrow$      & Tau $\uparrow$   & Top1-IR $\uparrow$          \\ 
\hline
SQL                                                         & \textbf{0.7360} & \textbf{0.4746} & \underline{0.1119}  & \textbf{0.2718}     & \textbf{0.3269}        & \textbf{0.2264}    \\
Plan                                                        & 0.7263 & 0.4680 & \textbf{0.1077}  & 0.2601     & 0.2490        & 0.1831    \\

Log  &  0.7088      &        0.4241       &      0.1401      &   0.2165   &    0.2216     &   0.1663         \\

KPI                                    &   0.7178    &    0.4552    &    0.1357     &    0.2296     &      0.2305      &   0.1721\\
\bottomrule
\end{tabular}
}
\label{table:mainModalStudyResult}
\end{table}

\subsubsection{Main Modality Study}
To investigate which data modality is most important for determining root cause impacts, we use \verb|SQL|, \verb|Plan|, \verb|Log|, and \verb|KPIs| as the main modalities, each fused with the other modalities. We evaluate the performance of using different modalities as the main modality on Hologres across all metrics.

Table~\ref{table:mainModalStudyResult} shows that using \verb|SQL| as the main modality yields the best performance, not only in $\textit{Top1-IR}$ but also across other metrics. The database instance information contained in the \verb|KPIs| modality plays an auxiliary role in determining root causes and their impact. Using KPIs that contain less information about the queries themselves as the main modality. The performance of using the \verb|Plan| modality as the main modality is very similar to using \verb|SQL| as the main modality because execution plans contain the columns involved in the queries and their execution orders. However, it is worth noting that execution plans in \verb|Plan| do not reflect the actual execution plan during query execution. 
\verb|SQL| benefits from using the pre-trained text encoder BERT, although the encoder is not specifically trained using SQL statements. However, table and column names in queries often contain semantic information, which can be captured by the text encoder, thereby reflecting the characteristics of the query.

\section{Conclusion}

This paper aims to identify root cause types and to rank them according to their impact on slow queries. We 
propose a multimodal Ranking framework for the Root Causes of slow queries (\texttt{RCRank}), which formulates diagnosis as a multimodal machine learning problem, leveraging multimodal information from query statements, execution plans, execution logs, and key performance indicators.
We propose a multimodal diagnosis model, which enables expressive embeddings of heterogeneous modalities with modal alignment and task relevance, effective and adaptive fusion of multimodal features, and unified identification and ranking of root causes. 
Experimental results show that our method is capable of outperforming existing methods regarding root cause identification and ranking. 
In future research, it is of interest to study how to best make a revision plan w.r.t. a set of slow queries, and study how to efficiently incrementally~\cite{haoicde24} train \texttt{RCRank} to include new root causes. 

\begin{acks}
 This work was partially supported by National Natural Science Foundation of China (62406112, 62372179) and Alibaba Innovative Research Program.
\end{acks}

%\clearpage

\bibliographystyle{ACM-Reference-Format}
\bibliography{sample}

\end{document}